\documentclass[%
 reprint,
 amsmath,amssymb,
 aps,
aps,prx,
]{revtex4-2} %linenumbers,

\usepackage{graphicx} % Include figure files
\usepackage{dcolumn}  % Align table columns on decimal point
\usepackage{bm}       % bold math
\usepackage{verbatim} %add comments
\usepackage{gensymb}
\usepackage{lineno}   % numbers of rows
\usepackage{soul}     %
\usepackage{color}
\usepackage{textcomp}
\usepackage{CJK}
\usepackage{multirow}
\usepackage[super]{nth}

\begin{document}
%\nolinenumbers

\preprint{APS/123-QED}

\title{ \textbf{}Single beam acoustical tweezers based on focused beams: A numerical analysis of 2D and 3D trapping capabilities.}% Force line breaks with \\
%Static holographic acoustical tweezers for dynamic manipulation: Focused vortex

\author{Zhixiong Gong}
%\email{zhixiong.gong@iemn.fr}
\affiliation{Univ. Lille, CNRS, Centrale Lille, Univ. Polytechnique Hauts-de-France, UMR 8520 -
IEMN - Institut d’Électronique de Microélectronique et de Nanotechnologie, F-59000 Lille, France}%
\author{Michael Baudoin}%
\email{Corresponding author: michael.baudoin@univ-lille.fr}
\homepage{\mbox{http://films-lab.univ-lille1.fr/michael}}
\affiliation{Univ. Lille, CNRS, Centrale Lille, Univ. Polytechnique Hauts-de-France, UMR 8520 -
IEMN - Institut d’Électronique de Microélectronique et de Nanotechnologie, F-59000
Lille, France}%
\affiliation{Institut Universitaire de France, 1 rue Descartes, 75005 Paris}%

\date{\today}% It is always \today, today,

\begin{abstract}
Selective single beam tweezers open tremendous perspectives in microfluidics and microbiology for the micromanipulation, assembly and mechanical properties testing of microparticles, cells and microorganisms. In optics, single beam optical tweezers rely on tightly focused laser beams, generating a three-dimensional (3D) trap at the focal point. In acoustics, 3D traps have so-far only been reported experimentally with specific wavefields called acoustical vortices. Indeed, many types of particles are expelled (not attracted to) the center of a focused beam. Yet the trapping capabilities of focused beams have so-far only been partially explored. In this paper, we explore numerically with an angular spectrum code the trapping capabilities of focused beams on a wide range of parameters (size over wavelength ratio and type of particles). We demonstrate (i) that 3D trapping of particles, droplets and microorganisms more compressible than the surrounding fluid is possible \textit{in and beyond Rayleigh regime} (e.g. polydimethylsiloxane, olive oil, benzene, and lipid sphere) and (ii) that 2D trapping (without axial trap) of particles with positive contrast factor can be  achieved by using the particles resonances.
\end{abstract}

\pacs{Valid PACS appear here}% PACS, the Physics and Astronomy
%\keywords{Suggested keywords}%Use showkeys class option if keyword display desired
\maketitle

%------------------------------------------------------------------------------------------------
\section{\label{sec:Introduction}Introduction}
3D microparticle trapping with single beam optical tweezers was demonstrated by Ashkin \textit{et al}. in 1986 with a focused laser beam under the condition that the particle refractive index is higher than that of the surrounding fluid medium \cite{ashkin1986observation}. The advantages of using a simple focused beam to manipulate particles include: (i) Simplicity: a focused beam is easy to produce with a simple lens. (ii) Excellent selectivity: the beam is focused on the target particle and hence has little effect on the neighbouring ones and (iii) Strong trap: A focused beam leads to strong gradients near the focal point that are thus suitable to create strong traps.

In acoustics the first to consider focused beams to trap object was Wu in 1991 \cite{wu1991acoustical}. But Wu used two collimated beams (not a single beam) propagating in opposite directions to obtain an acoustic trap for latex particles and clusters of frog eggs. By \textit{single beam} we mean a beam whose energy comes from only one direction of the space. Single beam tweezers are more convenient to use experimentally since they do not require to put sources or reflectors all around the target objects. Later on in 2009, as an analogy to optical tweezers, Shung's  group \cite{Shung2009single} explored the possibilities offered by single focused beams to trap particles. In their pionneering work, they succeeded to trap laterally (hence in 2D) oleic acid lipid droplets with a single beam at around 30 MHz in the Mie regime (with $a/\lambda \approx1.26$, where $a$ designates the particle radius and $\lambda$ the wavelength). However, axial trapping was not demonstrated. In addition, the ray method used to guide and analyse their experiments \cite{lee2005theoretical,lee2006radiation} was used beyond its limit of validity since such a method can only be used when $a \gg \lambda$. Later on, the same group demonstrated experimentally 2D trapping of a single elastic particle  and human cell beyond the Rayleigh regime with focused beams tweezers at frequencies up to 400 MHz  \cite{zheng2012acoustic,chen2017adjustable,lee2011targeted,liu2017single}. Note that the Rayleigh regime corresponds to the long wavelength regime wherein $a \ll \lambda$, while the Mie regime corresponds to $a \gtrsim \lambda$. They also performed calibration and measurement of sound forces on liquid droplets  \cite{lee2010calibration,lim2016calibration}. All these demonstration were however limited to 2D traps. Then, Silva \textit{et al.} \cite{silva2015computing} explored numerically with partial wave expansion the possibility to trap droplets in 3D  with focused beams. They showed theoretically that 3D trap can be obtained in the Rayleigh regime for specific silicone-oil droplet with a density of 1004 kg/m$^3$ closed to the one of water and a compressibility of 1050$\times 10^{-12}$ Pa, i.e. more than two times the one of water. However, (i) the trap was obtained only with droplets more compressible than the surrounding phase and (ii) 3D trapping beyond the Rayleigh regime was not explored. Finally, more recently Yang et al. \cite{cai2021self} reported the levitation upward of relatively large PDMS particle of radius 400 to 800 microns induced by a 1 MHz focused beam in the regime $a/\lambda \in [0.26 ,\, 0.52]$. Again 3D trap was not demonstrated in this regime.

Experimentally, 3D trapping against gravity of particles denser and stiffer than the surrounding phase with single beam has only been demonstrated with specific wavefields called focused acoustical vortices \cite{baresch2016observation}, some helical wave spinning around a phase singularity \cite{arfm_baudoin_2020}. Indeed, as we shall see in this paper, 3D particle trapping of these types of particles is not possible at the center of a focused beam. With acoustical vortices, 2D trapping of small microparticles \cite{baudoin2019folding, prap_sahely_2022} and cells \cite{baudoin2020spatially} has also been demonstrated with holographic tweezers based on spiraling interdigitated transducers \cite{prap_riaud_2017}. Later on, it was shown theoretically  \cite{gong2021Ztrap} that these types of particles could also be trapped in 3D with these type of devices. The 3D manipulation of bubbles with axial compensation of Archimedes force and radiation force with these wave structures was also demonstrated by Baresch et al. \cite{pnas_baresch_2020}.

Yet, (i) as discussed in the first paragraph, tweezers based on focused beams would have many advantages in terms of simplicity, selectivity and trapping force compared to their vortex counterparts and (ii) their trapping capabilities has only been partially explored. In this paper, we investigate numerically with an angular spectrum code \cite{sapozhnikov2013radiation,baudoin2020spatially,gong2020equivalence} the trapping capabilities of focused beam on a wide range of parameters (size and type of particles or droplets) in and beyond Rayleigh regime. We demonstrate (i) that 3D trapping is possible for some elastic and fluid particles more compressible than the surrounding phase [e.g.,  Polydimethylsiloxane (PDMS), olive oil and benzene] in Rayleigh regime and also at some specific frequencies beyond the Rayleigh regime, (ii) that only 2D lateral trapping of the most common elastic particles used in experiments (Pyrex, Polystyrene (PS) and Polyethylene (PE)) can be achieved at specific frequencies in the Mie regime, and (iii) that typical human cells cannot be trapped in 3D in a spherical focused beam, while the lipid (fat) cell 3D trapping is possible at some specific frequencies. This work is organized as follows: Section \ref{sec1:idt} describes how focused beams can be synthesized with holographic transducers. Section \ref{sec2:forcecalculation} A. and B. give an overview of the angular spectrum method and of the generalized Gor'kov theory  used in this paper to compute the radiation force for an arbitrary particle size and in the Rayleigh regime respectively. Section \ref{sec3: 3D trapping} A. and B. discusses the 2D and 3D trapping ability for two groups of particles, while biological particles are addressed in Sec. \ref{sec3C:3D cells trapping}.  

\section{\label{sec1:idt} Synthesis of focused beams with active holograms}

\begin{figure} [!htbp]
\includegraphics[width=0.4 \textwidth]{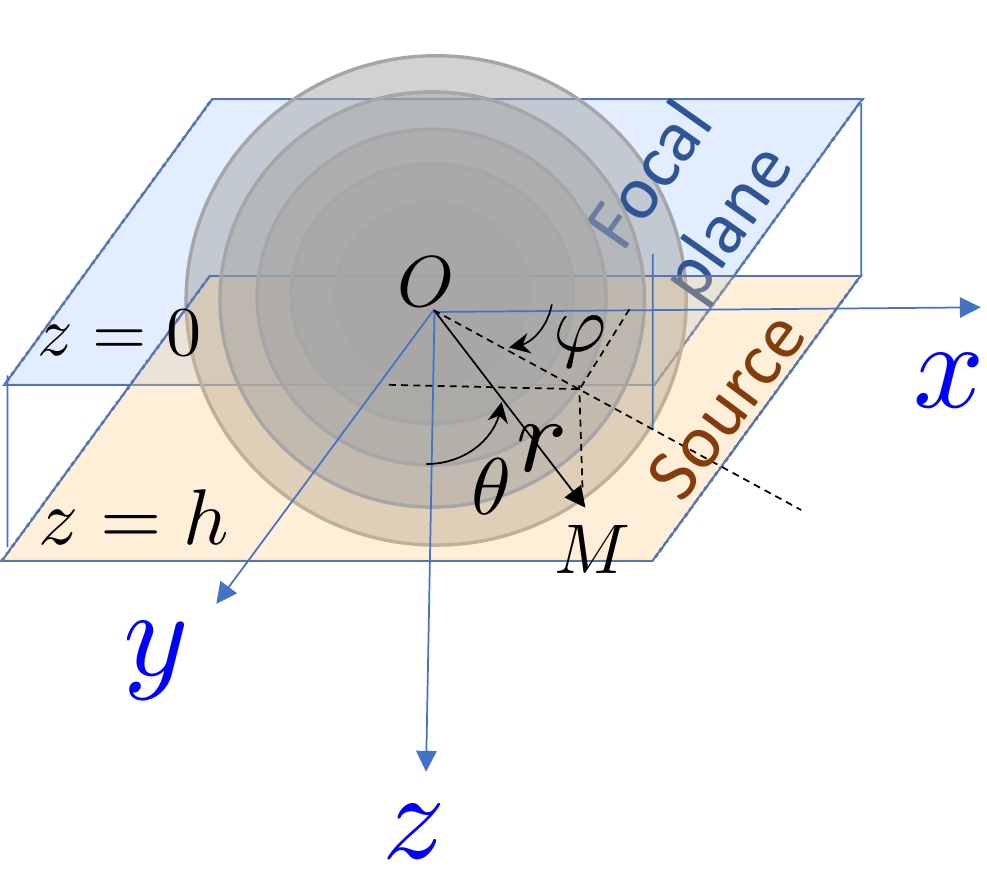} %0.34
\caption{Sketch representing the source and target plane and the spherical coordinates.}
\label{Fig:Sketch}
\end{figure}

It was shown recently by our team that complex high frequency acoustic fields such as focused acoustical vortices can be synthesized by using active holograms based on InterDigitated Transducers \cite{prap_riaud_2017,baudoin2019folding, arfm_baudoin_2020, prap_sahely_2022}. In short, the binarized phase hologram of the targeted wavefield is materialized by a set of metallic electrodes of inverse polarity deposited at the surface of a piezoelectric substrate. While generally holograms are passive and  require an external source, here the signal is directly synthesized by the electrode hologram which activate the piezoelectric substrate. The advantage of this method is that the transducers are flat, transparent, miniaturizable, and can produce high frequency signals to trap small particles and can be adapted to synthesize arbitrary wavefields.

The simulations of this paper are conducted with realistic fields, produced by binary phase holograms designed to produce focused beams at $40$ MHz (which corresponds to a wavelength of $\lambda \approx 37.5 \, \mu$m in water). The design of the electrodes are simply obtained by taking the intersection of a converging focused beam with a source plane (see Fig. \ref{Fig:Sketch}) and determining two set of equiphases lines in opposition of phase, which constitute a binary hologram of the targeted wavefield.  An ideal converging spherical focused beam is described in the spherical coordinates $(r,\theta, \varphi)$ as:
\begin{equation}
p^{*}(r, \theta, \varphi) = p_0 e^{i(kr - \omega t)}/r,
\label{eq:ideal Hankel vortex}
\end{equation}
where $p_0$ is the wave  amplitude, $k = \omega / c$ the wavenumber, $c$ the sound speed, and $t$ the time. The equiphase surfaces hence simply correspond to:
\begin{align}
\begin{split}
% first vortex
\phi &=\arg \left(p^{*}\right)= kr-\omega t = C + 2 n \pi
\label{eq:constan1}
\end{split}
\end{align}
where $C$ is a constant, $n$ an arbitrary integer and the time $t$ can be chosen arbitrarily and is hence chosen as $t=0$ in the following calculation. To obtain the intersection of these equiphase spherical surfaces with the source plane, we must introduce the cylindrical coordinates $(R, \varphi, z)$, take $z$ as a constant $z=h$ in equations \eqref{eq:constan1}, where $h$ is the distance between the source plane and the focal point (Fig. \ref{Fig:Sketch}). Since $R = \sqrt{r^2 - z^2}$, we simply obtain the following equations for the radius of the two set of electrodes of opposite phases:
\begin{eqnarray}
& & R_1 = \frac{1}{k} \sqrt{(C+2 n \pi)^2 - (kh)^2} \label{circ1} \\
& & R_2 = \frac{1}{k} \sqrt{C+ (2 n+1) \pi) - (kh)^2}  \label{circ2}
\end{eqnarray}
Note that the arbitrary constant $C$ must be chosen so that $C> kh$ and that each value of $n$ corresponds to a set of two electrodes. Hence the aperture of the transducer can be fixed by setting the maximum value $N$ of $n$, with $n \in [0,N]$. Here, $N$ was chosen to obtain an aperture angle of $\approx 60^{\circ}$ for a focal depth of $h = $1 mm.

\begin{figure*} [!htbp]
\includegraphics[width=17.6cm]{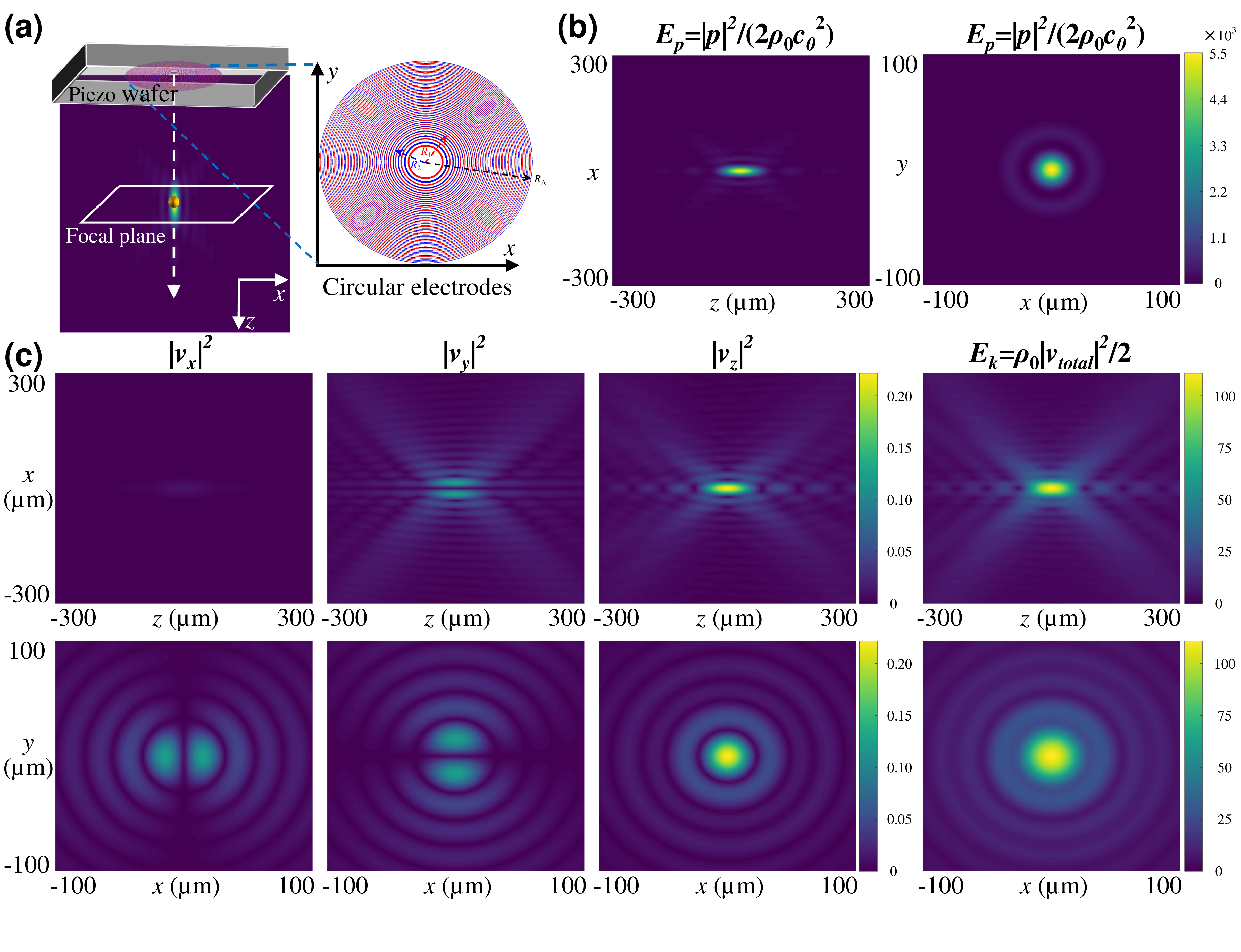}
\caption{(a) Schematic of a focused beam of finite aperture $60^o$ synthesized by a binary phase holograms representing the signal that would be generated by circular InterDigitated Transducers. The shape of the circular electrodes are given in the enlarged figure with the designed principle given in Sec. \ref{sec1:idt}.
(b) The acoustic potential energy $E_p = |p|^2/(2 \rho_0 c_0^2)$ in $(x,y=0,z)$ and $(x,y,z=0)$ planes. The pressure amplitude in the source plane is 0.1 MPa for all the simulations.
(c) The square amplitudes of the three components of the acoustic velocity $|\mathbf{v_{x,y,z}}|^2$ and the kinetic energy $E_k = \rho_0 |\mathbf{v}_{total}|^2/2$ are depicted in $(x,y,z=0)$ plane in the upper row, and in the $(x,y=0,z)$ plane in the lower row. The information are helpful to understand the trapping properties of Rayleigh particles (size much smaller than the wavelength) in 3D.}
\label{Fig1: Schematic and acoustic field}
\end{figure*}

The two set of electrodes are represented on Fig. \ref{Fig1: Schematic and acoustic field}(a) and (b). Compared to the binary holograms of focused acoustical vortices, the geometrical radii of the electrodes do not evolve with the azimuthal angle resulting in two set of circular concentric circles instead of spiraling ones. The focalization results from the decrease of the radial distance between two consecutive electrodes following the principle of Fresnel lenses. The width of the electrodes is chosen as half the distance between two consecutive electrodes of inverse polarity given by Eqs. \eqref{circ1} and \eqref{circ2}. The pressure and velocity fields produced by these transducers are calculated with an angular spectrum method \cite{sapozhnikov2013radiation,gong2020equivalence,baudoin2020spatially} and represented on Fig. \ref{Fig1: Schematic and acoustic field}. The angular spectrum simply consists in (i) taking the 2D Fourier Transform of the source to transform it into a sum of plane wave, (ii) propagating each plane wave up to the target plane and (iii) taking the inverse Fourier transform of the sum of the transported plane waves \cite{gong2021Ztrap}. 

\section{Calculation of the acoustic radiation force \label{sec2:forcecalculation}}

\subsection{General case}

The next step is to compute the radiation force that would be exerted on a particle depending on its position, size and composition. Different analytical formulations of the radiation force exerted by an arbitrary acoustic field on an arbitrary spherical particles have been derived by Silva \cite{jasa_silva_2011},  Baresch et al. \cite{jasa_baresch_2013} and Sapozhinkov \& Bailey  \cite{sapozhnikov2013radiation}, whose equivalence has been demonstrated by Gong \& Baudoin \cite{gong2020equivalence}. Note the acoustic radiation torque can also be calculated using the formulas proposed by \cite{epl_silva_2012} and \cite{gong2020acousticart}. Here we use an homemade code \cite{baudoin2020spatially} based on the angular spectrum method \cite{sapozhnikov2013radiation} to compute the acoustic radiation force, which is more direct considering that we also compute the acoustic field produced by the phase hologram with the angular spectrum method. Hence the force is calculated using the following formulas from Ref. \cite{gong2020equivalence,gong2021Ztrap}:

\begin{widetext}  % write long formulas in two-column
\begin{subequations}
\begin{eqnarray}
F_{x}&=&\frac{1}{4 \pi^{2} \rho_{0} k^{2} c^{2}}  \operatorname{Re} \left\{ \sum_{n=0}^{\infty} \sum_{m=-n}^{n} C_{n} \left(-b_{n+1}^{-m} 
H_{n m} H_{n+1, m-1}^*
+ b_{n+1}^{m} H_{n m} H_{n+1, m+1}^*
 \right) \right\}, \label{ASM_Fx}
\\
F_{y}&=&\frac{1}{4 \pi^{2} \rho_{0} k^{2} c^{2}} \operatorname{Im} \left\{ \sum_{n=0}^{\infty} \sum_{m=-n}^{n} C_{n} b_{n+1}^{m} \left(  H_{n,-m} H_{n+1, -m-1}^* +  H_{n m} H_{n+1, m+1}^* 
\right) \right\}, \label{ASM_Fy}
\\
F_{z}&=&-\frac{1}{2 \pi^{2} \rho_{0} k^{2} c^{2}} \operatorname{Re} \left\{ \sum_{n=0}^{\infty} \sum_{m=-n}^{n} C_{n} c_{n+1}^{m} 
H_{n m} H_{n+1, m}^* \right\} . \label{ASM_Fz}
\end{eqnarray}
\label{ASM force}
\end{subequations}
\end{widetext}
where $C_n = A_{n}+2 A_{n} A_{n+1}^*+A_{n+1}^*$, $b_{n}^{m}=\sqrt{[{(n+m)(n+m+1)}]/[{(2 n-1)(2 n+1)}]}$ and $c_{n}^{m}=\sqrt{[{(n+m)(n-m)}]/[{(2 n-1)(2 n+1)}]}$. Note that here the partial wave coefficients $A_{n}^m$ reduce to $A_n$ owing to the spherical shape of the particle.
The radiation force for general shapes can be obtained from Eq. (13) in Ref. \cite{gong2020equivalence} with $C_n^m$ and $C_n^{m \mp 1}$ given therein.

\subsection{Simplification in the Rayleigh regime}

When the particles are much smaller that the wavelength, i.e. in the Rayleigh regime, the radiation force formulas for spherical particles in Eq. \eqref{ASM force} simplify into \cite{sapozhnikov2013radiation}:

\begin{align}
\mathbf{F} =  & \, V_0 \left\{- \nabla \left[ f_1 \left( \frac{|p|^2}{4 \rho_0 c_0^2} \right) - f_2 \left( \frac{\rho_0 | \mathbf{v} |^2}{4} \right) \right]  \right.  \nonumber \\
& + \frac{(ka)^3}{3} \left[ \left( f_1^2 + \frac{2f_1 f_2}{3} \right)  
\mathrm{Re} \left( \frac{k}{c_0} p \mathbf{v}^* \right) \right.  \nonumber \\
&   \left. - \left. \frac{f_2^2}{3} \mathrm{Im} \left( \frac{\rho_0}{2} \mathbf{v}. \nabla \mathbf{v}^* \right)  \right]  \right\},
\label{Gorkov and scattering}
\end{align}
where $V_0 = 4/3 \, \pi a^3$ is the volume of the spherical shape with the particle radius $a$, $f_1$ and $f_2$ are the monopolar and dipolar acoustic contrast factors, $p$ and $\mathbf{v}$ are the complex pressure and velocity of the incident acoustic fields, ``Re" and ``Im'' designate respectively the real and imaginary part of a complex number, and the supercript ``$^*$" stands for the complex conjugate.
The first term in the curly braces of Eq. (\ref{Gorkov and scattering}) is nothing but the seminal Gor'kov expression of the radiation force introduced in \cite{gorkov1962forces}, which describes the contribution of a gradient force $\mathbf{F}_{\mbox{grad}} = -\nabla U$, with $U =  V_0 \left[ f_1 \left( |p|^2 / 4 \rho_0 c_0^2 \right) - f_2 \left( \rho_0 | \mathbf{v} |^2 / 4 \right) \right]$ the so-called Gor'kov potential. The remaining terms correspond to the scattering force $\mathbf{F}_{\mbox{scat}}$ contributions. Note that for a standing wave, the \nth{2} and \nth{3} terms vanish, while for a plane progressive wave, the Gor'kov gradient force vanish since $|p|^2$ and $| \mathbf{v} |^2$ are homogeneous. Also, as can been seen from this formula the gradient force is proportional to $O((ka)^3)$, while the scattering force is proportional to $O((ka)^6)$, so that the gradient forces are generally dominant (if they do not vanish) over the scattering force in the Rayleigh regime.

In the expression of the gradient force, the first term is proportional to the monopolar acoustic contrast factor $f_1 = (1 - \kappa_p / \kappa_0)$ and the acoustic potential energy density $E_p = |p|^2 / 2 \rho_0 c_0^2$, with $\kappa_p / \kappa_0$ the compressibility contrast between the particle and surrounding fluid. Hence this term is related to the relative compression/expansion of the particle compared to the surrounding fluid. Particles less compressible than the surrounding fluid $f_1 > 0$ (respectively more compressible, $f_1 < 0$) are hence pushed by this term toward the pressure amplitude minima (respectively maxima) of a wavefield in the Rayleigh regime. The second term of the gradient force is proportional to the dipolar acoustic contrast factor $f_2 = 3 (\rho_p - \rho_0) / (2 \rho_p + \rho_0)$ and the kinetic energy density $E_k = \rho_0 | \mathbf{v} |^2 / 2$, with $\rho_p$ / $\rho_0$ the density contrast between the particle and the surrounding fluid. Hence this term is related to the particle back and forth relative translation compared to the surrounding fluid. Particles denser than the surrounding fluid, i.e. with $f_2 > 0$ (respectively less dense,  $f_2 < 0$) are pushed by this term toward velocity amplitude maxima (respectively minima). If both the density and compressibility differ and for standing wavefields (wherein the pressure antinodes coincide to velocity nodes) it is convenient to introduce the so-called \textit{acoustic contrast factor} $\Phi_{SW}$ (\cite{laurell2010continuous,loc_bruus_2012}):
\begin{equation}
\Phi_{SW}=\frac{5 \rho_{p}-2 \rho_{0}}{2 \rho_{p}+\rho_{0}}-\frac{\kappa_{p}}{\kappa_{0}},
\label{eq: acoustical contrast factor}
\end{equation}
 whose sign indicates whether particle migrate toward pressure nodes or antinodes. Particle with positive contrast factor $\Phi_{SW} > 0$ (e.g., most solid particles and typical cells) get trapped at the pressure nodes, while particles with negative contrast factor $\Phi_{SW} < 0$, (e.g. certain liquids droplets) get trapped at the pressure antinodes.

\begin{table*}[!htbp]
\small
  \caption{ Acoustic properties for particles and fluid medium (water). Density $\rho_{0,p}$, Longitudinal speed of sound $c_l$, Shear speed of sound $c_t$, compressibility $\kappa = 1/K$ with modulus $K_{e}=\rho_{p}\left(c_{l}^{2}-4 / 3 c_{t}^{2}\right)$ for elastic material and $K_{f}=\rho_p c_l^{2}$ for fluid material. Compared with water, the Pyrex, Polystyrene (PS), Polyethylene (PE), typical human cells are less compressible, while Olive oil, Benzene, fat and Polydimethylsiloxane (PDMS) are more compressible. The acoustic contrast factor $\Phi_{SW}$ for Rayleigh particles in standing waves is also given for convenience.}
  \label{Table 1 Acoustic properties}
  \begin{tabular*}{1\textwidth}{@{\extracolsep{\fill}}l|cccrccc}
    \hline 
    Material & $\rho_{0,p}$ (kg/m$^3$) &  $c_l$ (m/s) &  $c_t$ (m/s) & $\kappa$ (1/TPa) & $f_1$ & $f_2$ & $\Phi_{SW}=f_1 + f_2$  \\
    \hline 
    Water      & 1000   & 1500   & ...  &444 &      &        &  \\
    \hline 
    Pyrex      & 2230   & 5640   & 3280 &25  &0.942 &0.676   &1.618 \\
    PS         & 1050   & 2350   & 1100 &243 &0.452 &0.048   &0.500 \\
    PE         & 1000   & 2400   & 1000 &225 &0.492 &0       &0.492 \\
    Cell       &[1000-1210] & ...& ...&[330-440] &... &...   & ...  \\
    Cell(avg)  & 1105   &1535    & ... &385  &0.136 &0.098   &0.234 \\
    \hline
    PDMS       & 1030   & 1030   &110   &929 &-1.091 &0.029  &-1.062\\
    Olive oil  &900     &1440    & ...  &535 &-0.206 &-0.107 &-0.313\\
    Benzene    &870     &1295    & ...  &685 &-0.542 &-0.142 &-0.684\\
    Lipid (fat)& 950    & 1450   & ...  &500 &-0.126 &-0.052 &-0.178\\
        \hline
  \end{tabular*}
\end{table*}

However, such analysis with the contrast factor is only valid for standing waves in the Rayleigh regime, since (i) for more complex wavefields the pressure maxima (minima) do not necessarily coincide with the velocity minima (maxima) (see e.g. \cite{gong2020three} for a discussion of particle trapping with spherical Bessel beams in the Rayleigh regime), (ii) in cases wherein the gradient forces are small (due to the homogeneity of the kinetic and potential energy density), the scattering force can also play a role, and finally (iii) the radiation force cannot be decomposed into gradient and scattering force beyond the Rayleigh regime. Hence, for complex wavefields (such as the one-sided focused beam considered here) in the Rayleigh regime, the contribution from the potential energy and the kinetic energy should be considered separately and added. This is why it is interesting to represent both the potential and kinetic energy (proportional to the pressure magnitude square and velocity magnitude square) as is done in Fig. \ref{Fig1: Schematic and acoustic field}. This figure shows that: (i) For this type of one-sided focused beams the focal point corresponds to both pressure and velocity amplitude maxima. (ii) The focusing magnitude, and hence gradients of the potential energy are stronger than the ones of the kinetic energy, suggesting that for equivalent monopolar and dipolar contrast factors, the monopolar term will play a larger role than the dipolar term. 

\begin{figure} [!htbp]
\includegraphics[width=1\linewidth]{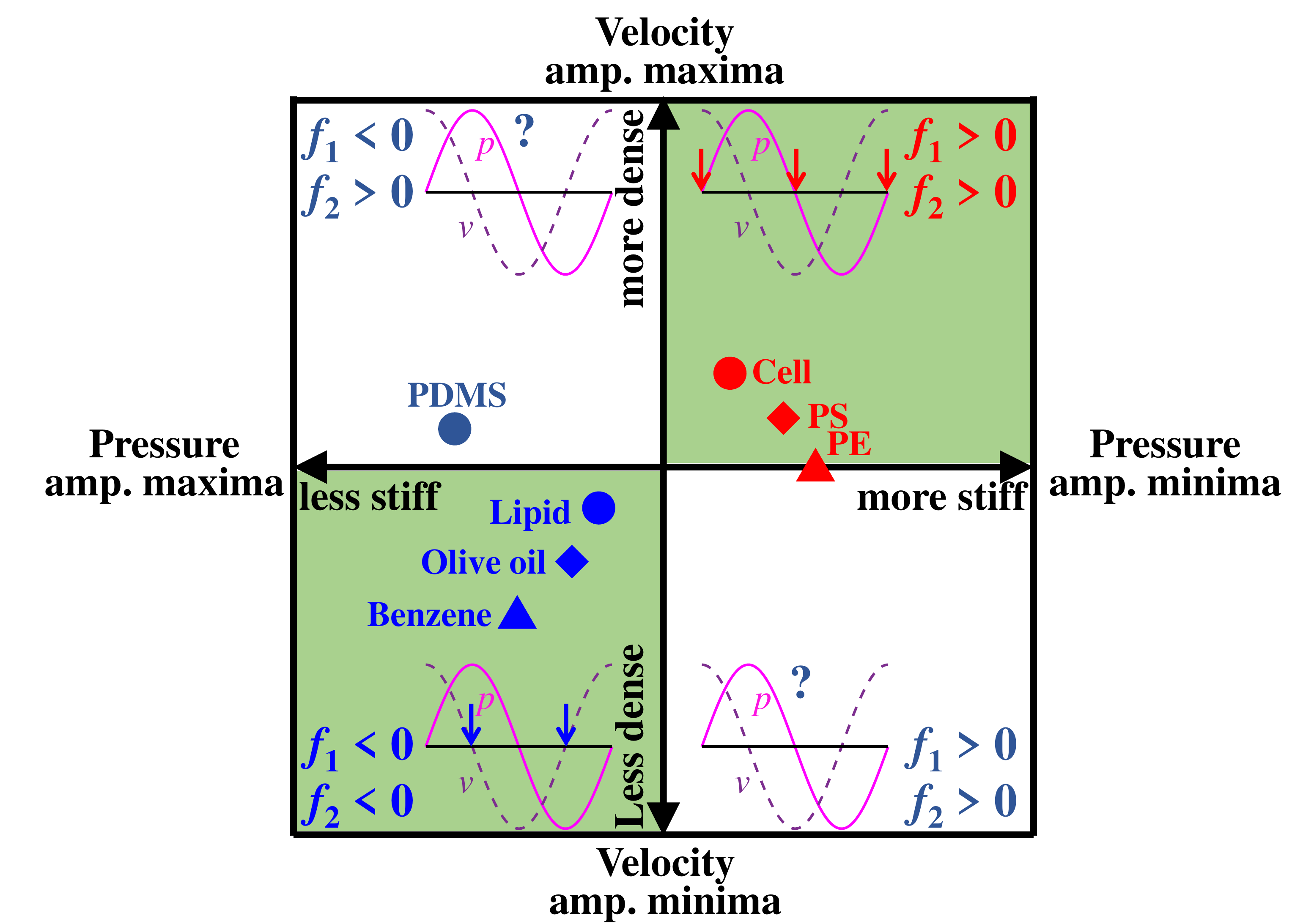}
\caption{Quadrant chart explaining in which direction (toward pressure/velocity field maxima or minima) are pushed different types of particles and cells by the monopole and dipole forces depending on the sign of the monopolar and dipolar contrast factors $f_1$ and $f_2$. The referenced medium is water with the acoustic parameters given in Table \ref{Table 1 Acoustic properties}. For a 1D standing wave, pressure maxima (anti-nodes) correspond to velocity minima (nodes). Hence the movement of the particles in green quadrants are obvious since both the kinetic and potential forces push in the same directions. In the other quadrants it is necessary to calculate the contrast factor $\Phi_{SW}$ to determine whether particle migrate to the nodes or antinodes. However, the motion is less obvious for focused beams wherein pressure maxima do not correspond to velocity minima and conversely.}
\label{Fig0: Schematic in a standing waves}
\end{figure}

In this paper, we consider many different types of particles and droplets suspended in water, whose properties are summarized in table \ref{Table 1 Acoustic properties}, insonified by the one-sided focused beam represented on Fig. \ref{Fig1: Schematic and acoustic field}. Fig. \ref{Fig0: Schematic in a standing waves} summarizes the expected role played by the monopolar (potential energy) and dipolar (kinetic energy) terms on the particle.

%-----Sec III
\section{\label{sec3: 3D trapping} 3D trapping with a focused-beam acoustical tweezers.}
In this section, we will study two groups of typical particles, which are commonly used in trapping experiments: one group [Pyrex, Polystyrene (PS), and Polyethylene (PE)] consists of materials less compressible ($f_1 >0$) and denser ($f_2 > 0$) than the surrounding medium, the other [Olive oil, Benzene, and Polydimethylsiloxane (PDMS)] uses materials more compressible ($f_1 < 0$) and generally lighter ($f_2 < 0$), except for PDMS, whose density is slightly higher than that of water. The detailed acoustic parameters used for the simulations are listed in Table \ref{Table 1 Acoustic properties}.

%-----------------------
\subsection{\label{sec3A: denser particle}Materials less compressible and denser than the surrounding medium}

The most common elastic particles used experimentally, namely Pyrex, Polystyrene (PS), and PolyEthylene (PE) belong to this category, as well as most typical human cells  \cite{augustsson2016iso}. The case of biological particles will be treated separately in subsection \ref{sec3C:3D cells trapping}. 

\subsubsection{\label{secA1: in Rayleigh regime}Radiation forces in Rayleigh regime}

\begin{figure*} [!htbp]
\includegraphics[width=17.6cm]{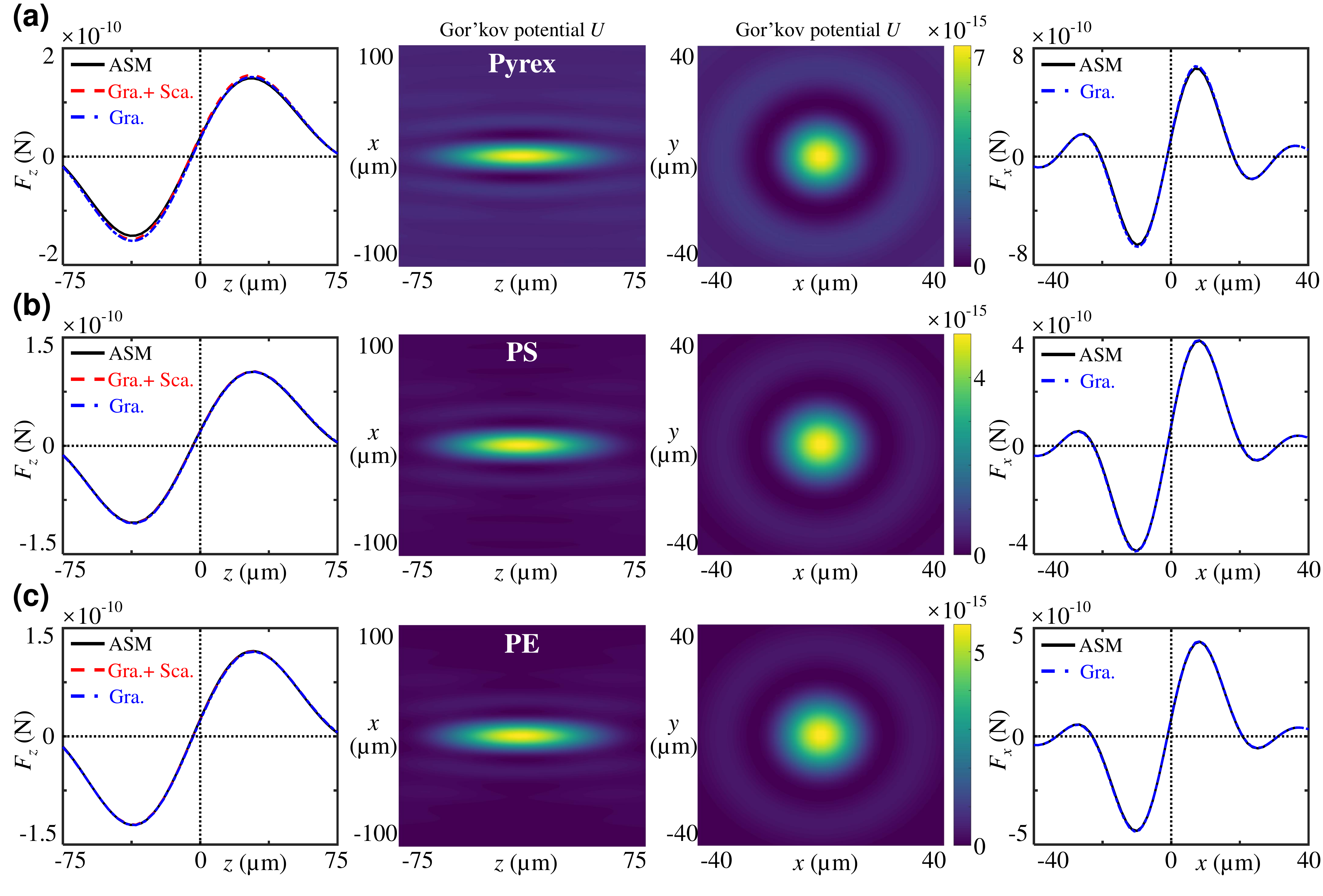}
\caption{Axial radiation forces $F_{z}$ (first column), Gor'kov potential $U$ (second and third columns) and lateral radiation force $F_{x}$ versus spatial positions for (a) Pyrex, (b) Polystyrene (PS), and (c) Polyethylene (PE) particles with the radius $a = 1 \, \mu$m. The Gor'kov potential $U$ exhibits a maximum in the beam center indicating that the gradient force $\mathbf{F}_{grad} = - \nabla U$ pushes the particle away from the focal point. The axial and lateral radiation forces versus their positions are given in the first and forth columns, respectively. Around the focal point ($x=y=z=0$) the force is negative (for $z<0$ and $x<0$)  and then positive (for $z>0$ and $x>0$) indicating that this point is repulsive for the particle.}
\label{Fig2: denser particle with radiaiton force and Gor'kov potential.}
\end{figure*}

In this section we consider microparticles of $1 \, \mu m$ in radius made of three different kinds of materials (Pyrex, PS, and PE) and insonified by the acoustic field introduced in Sec. \ref{sec1:idt}, with a wavelength over particle size ratio $a / \lambda \approx 0.03 \ll 1$. The axial radiation force $F_z$ and lateral radiation force $F_x$ (represented in Fig. \ref{Fig2: denser particle with radiaiton force and Gor'kov potential.}, first and last columns respectively) are calculated with three different methods: (i) with Gor'kov's original expression of the gradient force, (ii) with the generalized Gor'kov theory \cite{sapozhnikov2013radiation} taking into account the scattering contribution (Eq. \eqref{Gorkov and scattering}) and (iii) with the ASM complete expression of the force. As expected, the three calculations give similar results since (i) the calculation is performed in the Rayleigh regime $ka \ll 1$ and (ii) the field is not homogeneous and hence the contribution of the gradient force dominates over the one of the scattering force.  The Gor'kov potential $U$ is represented on the columns 2 and 3 of Fig. \ref{Fig2: denser particle with radiaiton force and Gor'kov potential.}. Stable positions correspond to minima of Gor'kov potential, while unstable ones to maxima of Gor'kov potential.

All these figures show that these types of particles are expelled both laterally and axially from the center of the focused beam. Indeed, the center of the focused beam is both a pressure and velocity magnitude maximum. Since $f_1 >0$ and $f_2>0$, the monopolar (potential energy) and the dipolar (kinetic energy) contributions of the gradient force will push respectively the particle \textit{away} from the beam center and \textit{toward} the beam center respectively. But since (i) the compressibility contrast is larger than the density contrast for these particles leading to $f_1 > f_2 $, and (ii) the gradients of the potential energy are stronger than the gradient of the kinetic energy (as discussed previously), the monopolar force is dominant and pushes the particles away from the beam center.

\begin{figure*} [!htbp]
\includegraphics[width=\linewidth]{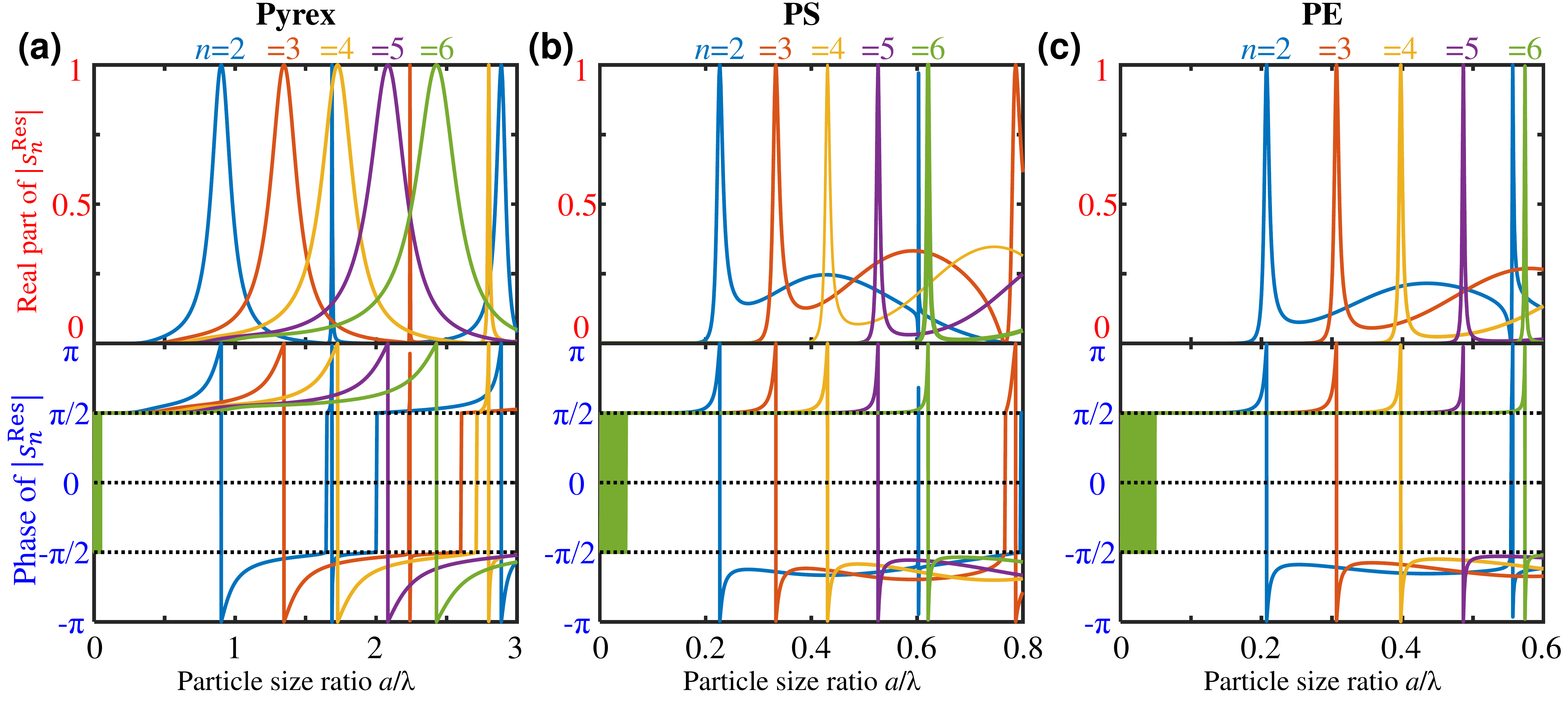}
\caption{First resonances of (a) Pyrex, (b) PS, and (c) PE particles. The first row gives the quantity of the real part of the resonance scattering coefficients $S_{Res}$, while the second gives the phase information for particle size ratio $a/\lambda$ from 0 to 3. It is clearly observed that there is a $\pi$ shift (see the below row) through the resonance peaks (see the upper row). Because the real part is close to 0, the phase has a fluctuations $\pi$ (the imaginary part is position) and $-\pi$ (negative). The explicit size ratio for different resonances are given in Table \ref{table 2 RST for denser sphere}.}
\label{Fig3: RST denser particles}
\end{figure*}
%-----------------------
\subsubsection{\label{secA2: RST theory}Resonance scattering theory}
Spherical particles constitute some cavities for the wave. Hence, the particles exhibit some resonances when the wavelength of the particle  approaches the size of the particle, leading to directional scattering patterns. These resonances play a fundamental role on the radiation force beyond Rayleigh regime. Hence, it is vital to determine the resonance frequencies of the particle depending on their composition to analyse the results. The resonance scattering theory developed by Flax \textit{et al.} has been widely used in the literature to isolate the  scattering resonances and can predict the correct magnitude of the scattering coefficients. However, they cannot predict the useful phase information \cite{flax1981theory}. This theory was further improved by Rhee \& Park \cite{rhee1997novel}, who derived expressions of both the magnitude and phase with the following resonance scattering coefficients:
\begin{equation}
s_n^{\text{Res}}=\frac{A_{n}-A_{n}^{b}}{1+2 A_{n}^{b}}, 
\label{Rigid background}
\end{equation}
where $A_n = (s_n-1)/2$ and $A_n^b$ are the total and background partial wave coefficients respectively. For elastic materials, a background scattering from a rigid particle with the same size can be used with the scattering coefficients given in Appendix \ref{Appendix A1} to only keep the contribution of the particle resonance. The scattering coefficients for the total (elastic resonance and rigid background) scattering are given in Appendix \ref{Appendix A4}. This method works quite well for dense metal materials, for example, the tungsten carbide with very sharp peaks of resonance \cite{gong2019resonance}. The improved method by Rhee \& Park is applied for the three materials in the present work. The real part and phase of the resonance scattering coefficients $s_n^{\text{Res}}$ with different orders $n \in [2,6]$ as a function of the particle radius over wavelength ratio $a/\lambda$ are represented in Fig. \ref{Fig3: RST denser particles} for (a) Pyrex, (b) PS, and (c) PE, respectively. The sharp peaks of real part of $s_n^{\text{Res}}$ with the corresponding $\pi$ shift in the phase information provide reliable resonances of orders $n=2, 3, 4, 5$ and $6$. The exact value of the radius over wavelength ratio for each resonance are collected in Table \ref{table 2 RST for denser sphere}.
\begin{table}[!htbp]
%\small
  \caption{Particle size ratios ($a/\lambda$) at resonance frequencies of different orders for Pyrex, PS and PE materials.}
  \label{table 2 RST for denser sphere}
  \setlength{\tabcolsep}{3mm}
  \begin{tabular}{l| c c c c c}
\hline
 Order &  $n=2$ & $n=3$ & $n=4$ & $n=5$ & $n=6$\\
\hline 
Pyrex &0.902 &1.347 & 1.727 & 2.084 & 2.428\\ 

PS &0.227 &0.333 & 0.431 & 0.526 & 0.621\\ 

PE &0.207 &0.306 & 0.398 & 0.486 &0.574 \\ 

 \hline
  \end{tabular}
\end{table}
%-----------------------
\begin{figure*} [!htbp]
\includegraphics[width=\linewidth]{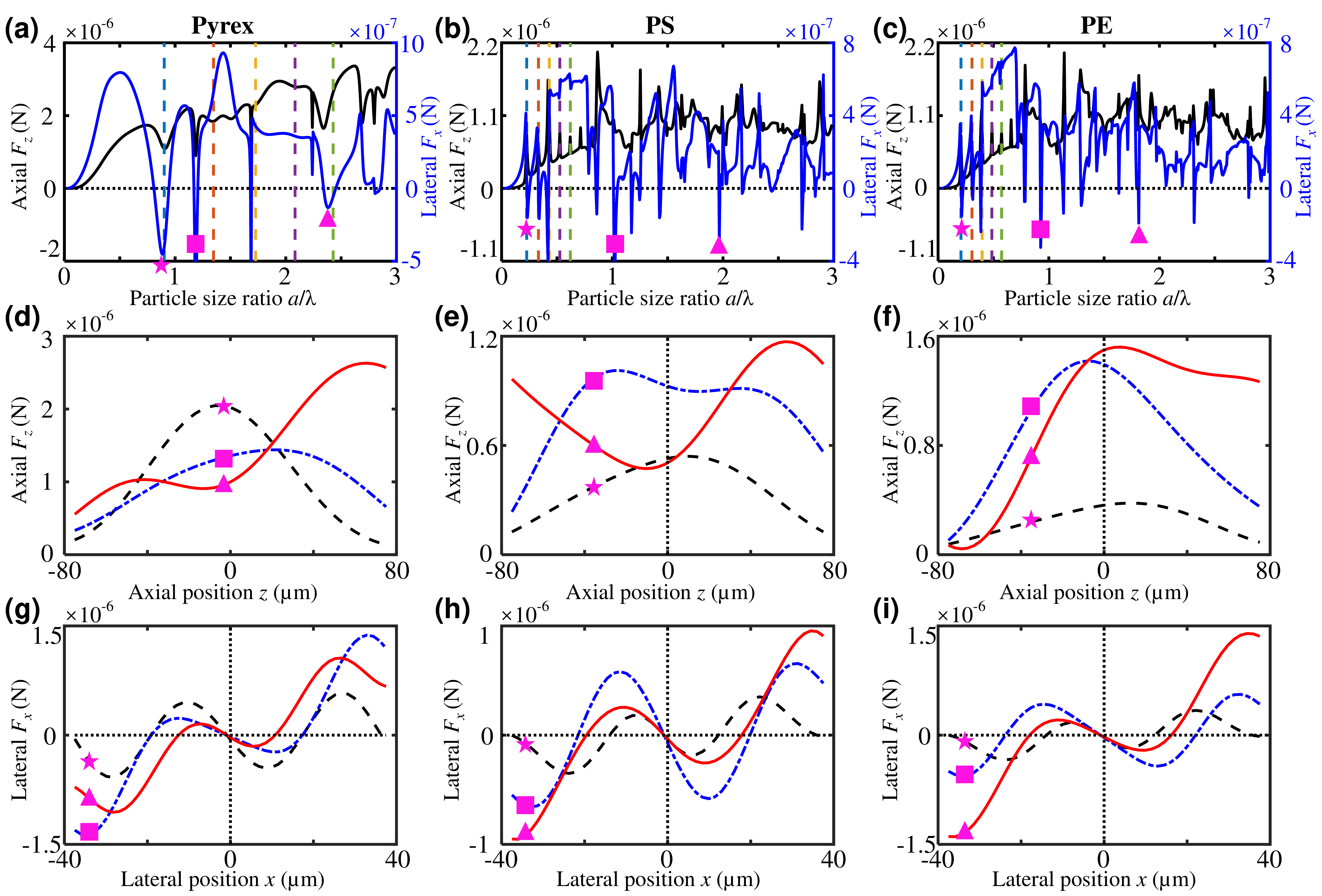}
\caption{Three-dimensional acoustical radiation forces based on the angular spectrum method at a fixed axial ($z_s = 30 \, \mu$m ) and lateral ($x_s = 8 \, \mu$m ) position for different particle materials: (a) Pyrex, (b) PS, and (c) PE. The left and right vertical axes are respective the axial ($F_z$, black solid line) and lateral ($F_x$, blue solid line) radiation forces, while the horizontal axis is the ratio of particle radius over the wavelength $a/\lambda$ from 0 to 3.
Three explicit size ratios are chosen to show the possibility of two dimensional trapping, whereas there is no axial trapping beyond the Rayleigh limit for (d,g) Pyrex with $a/\lambda =$ 0.88 (\textcolor{magenta}{$\bigstar$}), 1.20 (\textcolor{magenta}{$\blacksquare$}), and 2.37 (\textcolor{magenta}{$\blacktriangle$}); (e,h) PS with $a/\lambda =$ 0.23 (\textcolor{magenta}{$\bigstar$}), 1.02 (\textcolor{magenta}{$\blacksquare$}), and 1.97 (\textcolor{magenta}{$\blacktriangle$}); and (f,i) PE with $a/\lambda =$ 0.21 (\textcolor{magenta}{$\bigstar$}), 0.93 (\textcolor{magenta}{$\blacksquare$}), and 1.81 (\textcolor{magenta}{$\blacktriangle$}). It is possible to have lateral trapping but not axial trapping beyond the Rayleigh regime. While there is no axial or lateral trap in Rayleigh regime as shown in Fig. \ref{Fig2: denser particle with radiaiton force and Gor'kov potential.}.}
\label{Fig4: ASM denser particles}
\end{figure*}

%-----------------------
\subsubsection{\label{secA3: 3D trapping}3D radiation forces beyond Rayleigh regime and 2D trapping}
As shown in Sec. \ref{secA1: in Rayleigh regime} for the considered particles, there is no trapping in neither the axial nor the lateral directions in the Rayleigh regime. Here, we study the trapping possibility for these particles beyond Rayleigh regime based on the angular spectrum method. To obtain a 3D trap, restoring forces (pushing the particle toward the beam center) are required in both axial and lateral directions, which means that the radiation forces should be negative when the particle is slightly displaced along $z$ or $x$ axis ($z>0$, $x>0$) and positive when the particle is slightly displaced in the other directions ($z<0$, $x<0$). Of course the magnitude of the restoring force depends on the exact location of the particle. For simplicity and efficiency, we first calculate the axial and lateral radiation force as a function of the particle size ratios $a/\lambda = [0,3]$ at the fixed positions $(x_s,z_s) = (0,30) \,\mu$m and $(x_s,z_s) = (8,0) \, \mu$m respectively (see Fig. \ref{Fig4: ASM denser particles} (a,b,c)). These positions correspond to the maximum of the trapping force in the Rayleigh regime and remain in the central bright spot of the focused wave. An axial trap and a lateral trap at the center of the focused beam can be obtained only if the values of the axial and lateral forces at these positions are negative. Note that this is a necessary but not sufficient condition to obtain a trap. Fig. \ref{Fig4: ASM denser particles} shows that for Pyrex, PS and PE particles, the axial force (black curve) is always positive at these positions, hence precluding any possibility for axial trapping of these particles with focused waves. However, close to the particle resonances (calculated in the previous section and represented on these graphs by the dashed curves for n=[2,6]), there are some strong variation of the lateral force, which can become negative, hence suggesting that there is some lateral restoring force. To further demonstrate the 2D lateral trapping ability, three typical sizes of each particle are selected which have positive axial forces and negative lateral forces (corresponding to the star, rectangle and triangles marks on the graphs (a,b,c). The axial radiation forces versus $z$ on the beam axis are plotted in the second row (d,e,f), and the lateral forces versus $x$ at the focal plane ($z=0$) are given in the third row (g,h,i). It is clearly shown that the axial radiation forces is always positive, which will push the particles outside of the focus, hence confirming that there is no axial trapping whatever the position of the particle. The lateral force however is positive for $x<0$ and positive for $x>0$ confirming that the lateral trapping ability at these specific ratios of particle size over wavelength.

%-----------------------

\subsubsection{Conclusion}

Our calculations show that Pyrex, PS and PE particles are always expelled from the focal point in the Rayleigh regime. Beyond Rayleigh regime, only lateral trapping of these particles is possible at some specific particle size over wavelength ratios, close to some of the particle resonances. The particle is however always pushed in the axial direction by the radiation force.

\subsection{\label{sec3B: less dense} Materials more compressible and/or less dense than the surrounding medium}

\begin{figure*} [!htbp]
\includegraphics[width=17.6cm]{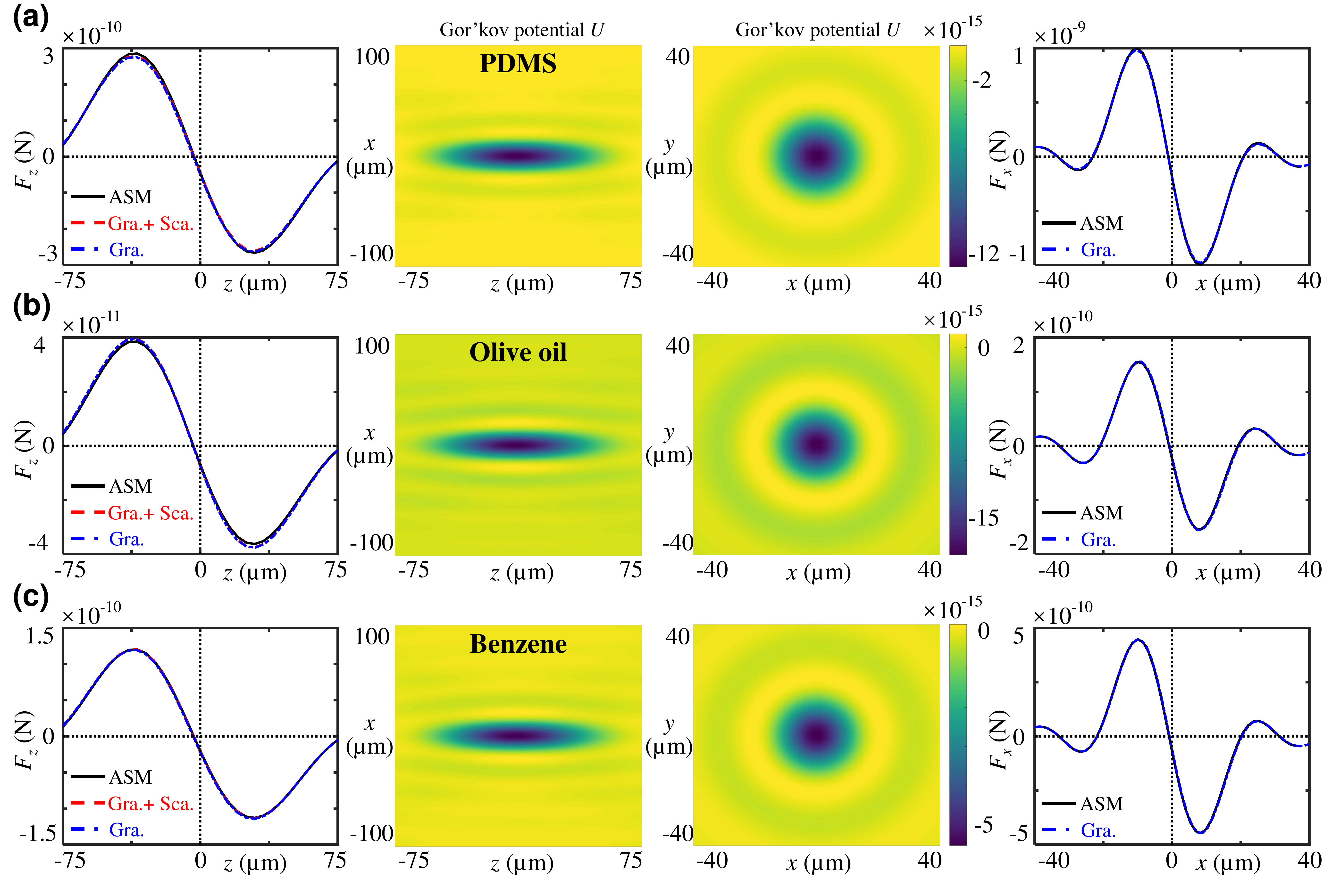}
\caption{Axial radiation forces $F_{z}$ (first column), Gor'kov potential $U$ (second and third columns) and lateral radiation force $F_{x}$ versus spatial positions for (a) Polydimethylsiloxane (PDMS), (b) Olive oil, (c) Benzene with the particle radius $a = 1$ $\mu$m. The Gor'kov potential $U$ exhibits a minimum in the beam center indicating that the gradient force $\mathbf{F}_{grad} = - \nabla U$ pushes the particle toward the focal point. The axial and lateral radiation forces versus their positions are given in the first and forth columns, respectively. The results show clearly a 3D trap at the center of the focused beam for the considered materials.}
\label{Fig5: light particle with radiaiton force and Gor'kov potential.}
\end{figure*}

The second group of materials (Olive oil, Benzene, Polydimethylsiloxane (PDMS)) considered in this section are more compressible ($f_1 < 0$) and generally lighter ($f_2 < 0$) than the surrounding water, except from PDMS which is slightly heavier than water.
%-----------------------
\subsubsection{\label{secB1: in Rayleigh regime}Radiation force in Rayleigh regime}

The same analysis as in section \ref{secA1: in Rayleigh regime} is conducted here for this new group of materials. Again the gradient force dominates, so that Gor'kov expression is sufficient to estimate the force in the Rayleigh regime. The results represented on Fig. \ref{Fig5: light particle with radiaiton force and Gor'kov potential.} show that all three types of particles are trapped in 3D at the focal point of focused beams in the Rayleigh regime. Indeed, PDMS is the perfect solid particle candidate to be trapped by a focused beam in the Rayleigh regime since it is both more compressible and denser than water, leading to $f_1 <0$ and $f_2 >0$ and hence to the contributions of both the potential and kinetic energy to the gradient force pushing the particle toward the focal point. For olive oil and benzene, $f_1$ and $f_2$ are negative. Hence the gradient of the potential energy pushes the droplet toward the focal point while the gradient of the kinetic energy pushes the droplet away from the focal point. But since $|f_1| > |f_2|$ and the gradient of the potential energy is stronger than the gradient of the kinetic energy, the contribution of the monopolar term to the gradient force dominates and hence the droplet is pushed toward the focal point, leading to 3D trapping of the particle. This is clearly seen on the representation of the Gor'kov potential $U$ (columns 2 and 3 of Fig. \ref{Fig5: light particle with radiaiton force and Gor'kov potential.}), which is minimum at the focal point and from the calculation of the axial and lateral forces (columns 1 and 4 from Fig. \ref{Fig5: light particle with radiaiton force and Gor'kov potential.}) which are positive and then negative around the focal point leading to some restoring force which make the particle converge toward the beam center.

%-----------------------
\subsubsection{\label{secB2: RST theory}Resonance scattering theory}

\begin{figure} [!htbp]
\includegraphics[width=0.8\linewidth]{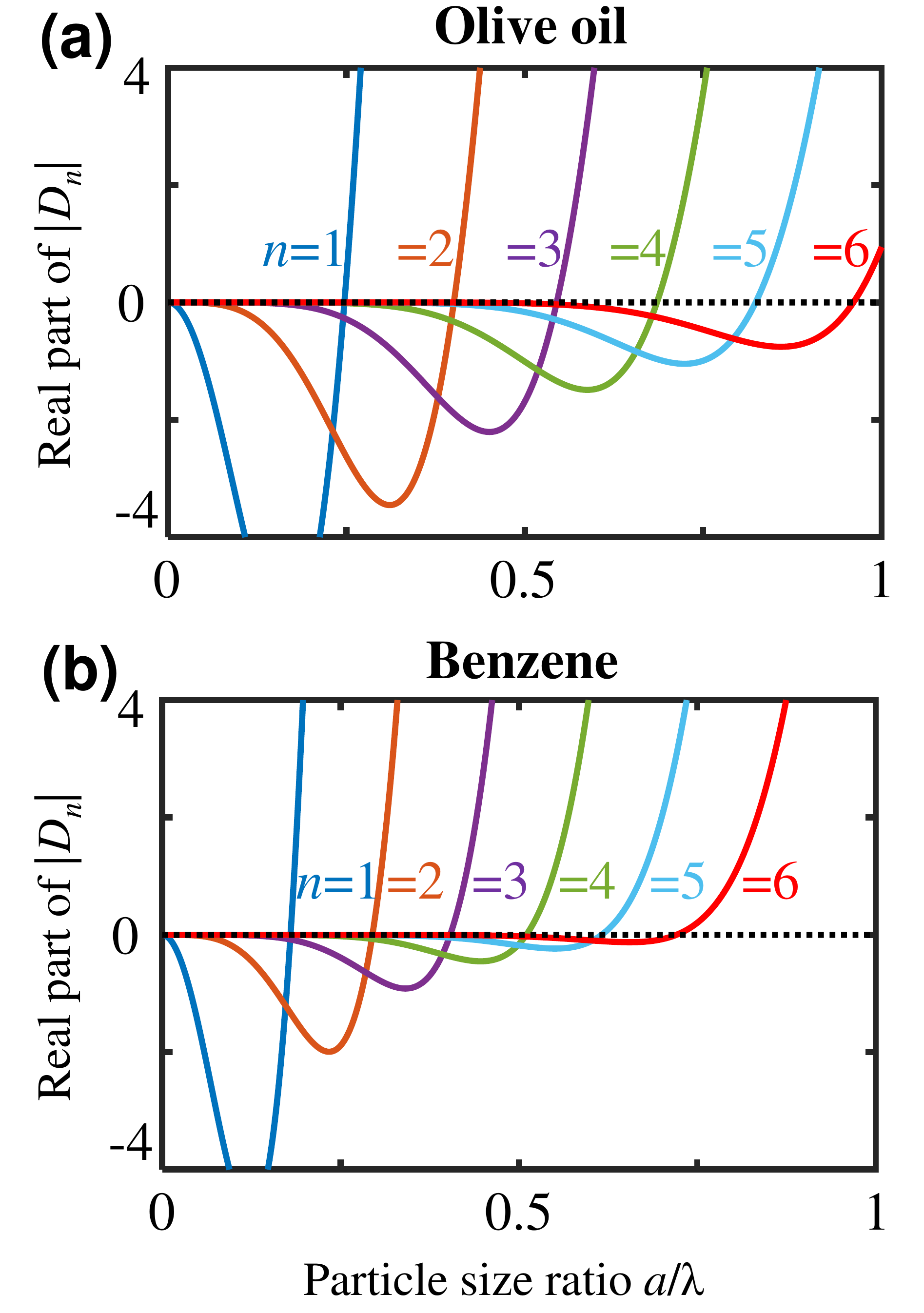}
\caption{ The resonance frequencies are clarified by finding out the zero value of the real part of $\left| D_n \right|$ with different order $n$ for fluid particles with materials (a) Olive oil and (b) Benzene. The explicit expression of $\left| D_n \right|$ for a fluid sphere is given in Appendix \ref{fluid sphere}. The size ratio of different resonances are given in Table \ref{Table 3 RST for fluid sphere}.}
\label{Fig6: RST less dense particles}
\end{figure}

Again it is interesting to localize the particle and droplet resonances before studying their ability to get trapped by a focused beam beyond the Rayleigh regime. Since the acoustic impedance of the considered particles and water are close, as shown in Table \ref{Table 1 Acoustic properties}, it is not easy to isolate the resonance contribution by subtracting the background from the total scattering field. For a fluid bubble, the scattering from a soft sphere could be taken as the background (see Appendix \ref{Appendix A2}). However, the soft background is not suitable for the liquid spheres whose density and velocity are close to those of the surrounding water. A more complicated intermediate (hybrid) background may be used \cite{gaunaurd1991sound,gong2019resonance} but this is still an open question. In this work, an alternative method is applied to clarify the resonance sizes for fluid spheres by finding out the real roots of $\text{Re}(D_n)=0$ for different $n$th partial wave \cite{marston1988gtd}, with $D_n$ a parameter introduced for convenience with the relation to the scattering coefficients $s_n=-D_n^{*}/D_n$ given in the Appendix \ref{Appendix A3}. The first several roots are clarified in Fig. \ref{Fig6: RST less dense particles} for fluid spheres of materials (a) olive oil and (b) benzene. The explicit resonance size ratios are listed in Table \ref{Table 3 RST for fluid sphere}. Note that the resonance sizes of PDMS spheres are not easy to determine and not discussed here.

\begin{table}[!htbp]
\small
  \caption{Particle sizes ($a/\lambda$) at resonance frequencies for Olive oil and Benzene.}
  \label{Table 3 RST for fluid sphere}
  \setlength{\tabcolsep}{2mm}
  \begin{tabular}{l| c c c c c c}
\hline
 Order &  $n=1$ & $n=2$ & $n=3$ & $n=4$ & $n=5$ & $n=6$\\
\hline 
%Pyrex &0.902 &1.347 & 1.727 & 2.084 & 2.428\\ 

Olive oil &0.247 &0.400 &0.545  &0.685  &0.824 &0.961 \\ 

Benzene &0.180 &0.295 &0.404  &0.510 &0.616 &0.721 \\ 

 \hline
  \end{tabular}
\end{table}

%-----------------------
\begin{figure*} [!htbp]
\includegraphics[width=17.6cm]{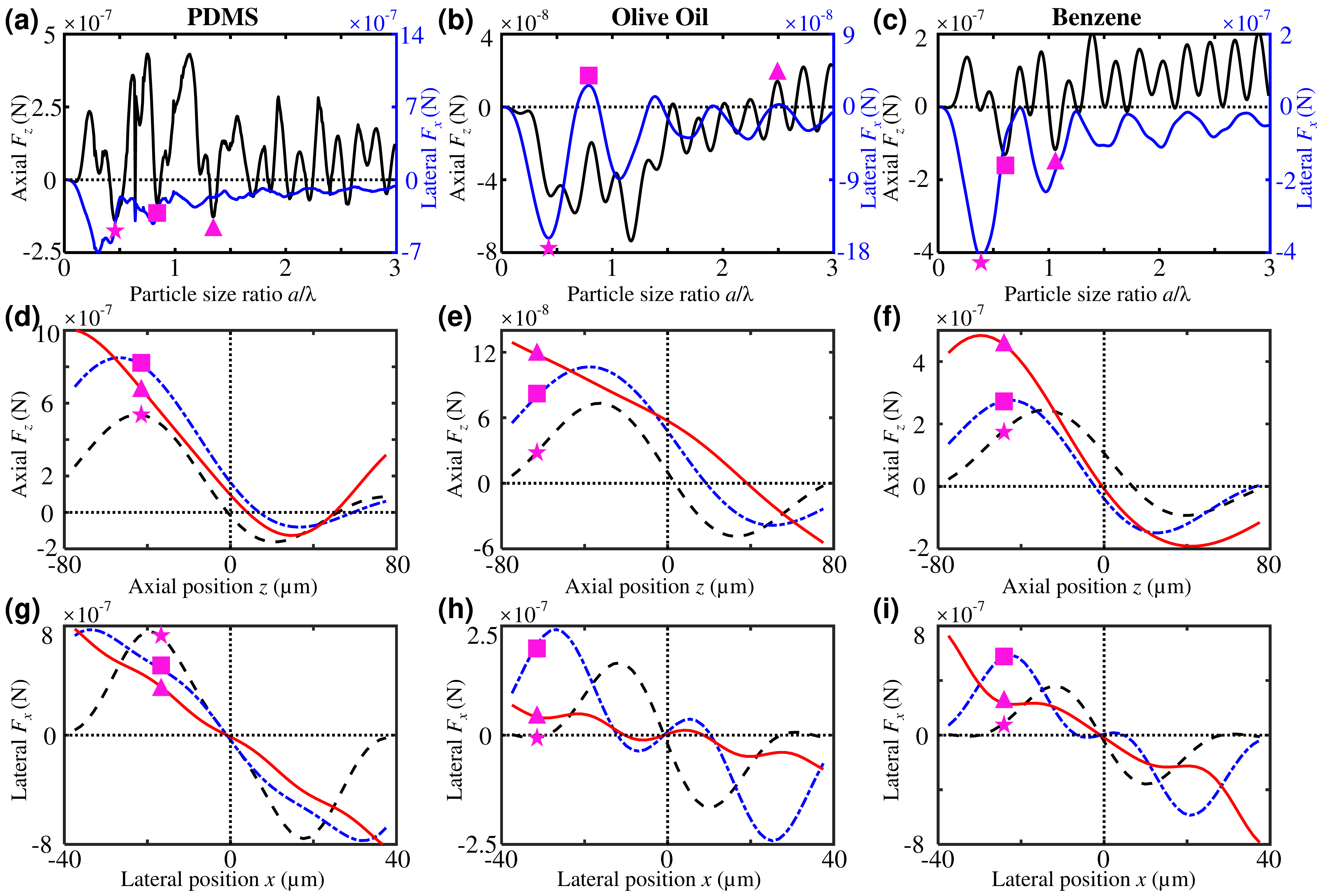}
\caption{Three-dimensional acoustical radiation forces based on the angular spectrum method at a fixed axial ($z_s = 30 \, \mu$m ) and lateral ($x_s = 8 \, \mu$m ) position for different particle materials: (a) PDMS, (b) Olive oil, and (c) Benzene. 
The left and right vertical axes are respective the axial ($F_z$, black solid line) and lateral ($F_x$, blue solid line) radiation force, while the horizontal axis is the ratio of particle radius over the wavelength $a/\lambda$ from 0 to 3.
Three explicit size ratios are chosen to show the possibility of three dimensional trapping beyond the Rayleigh limit for (d,g) PDMS with with $a/\lambda =$ 0.46 (\textcolor{magenta}{$\bigstar$}), 0.83 (\textcolor{magenta}{$\blacksquare$}), and 1.35 (\textcolor{magenta}{$\blacktriangle$}); 
(e,h) Olive oil with $a/\lambda =$ 0.43 (\textcolor{magenta}{$\bigstar$}), 0.79 (\textcolor{magenta}{$\blacksquare$}), and 2.49 (\textcolor{magenta}{$\blacktriangle$}); 
and (f,i) Benzene with $a/\lambda =$ 0.39 (\textcolor{magenta}{$\bigstar$}), 0.61 (\textcolor{magenta}{$\blacksquare$}), and 1.06 (\textcolor{magenta}{$\blacktriangle$}). The lateral radiation force versus $x$ in the third row are plotted at the axial equilibrium positions as obtained in the second. 3D trapping is possible for several particle sizes beyond Rayleigh regime.}
\label{Fig7: ASM less dense particles}
\end{figure*}

%-----------------------
\subsubsection{\label{secB3: 3D trapping}3D radiation forces beyond Rayleigh regime and 3D trapping}

The angular spectrum method is applied to compute the 3D radiation forces for the more compressible particles beyond the Rayleigh regime. The axial and lateral forces versus the particle size ratios $a/\lambda$ are given in Fig. \ref{Fig7: ASM less dense particles}(a-c). Similarly to Fig. \ref{Fig4: ASM denser particles}, the radiation force is first calculated at the axial fixed position $z_s = 30 \, \mu$m on the beam axis and the lateral fixed position is $x_s = 8 \, \mu$m in the focal plane. This time, the results show that both axial and lateral negative radiation forces can be observed for some  specific size over wavelength ratios, hence suggesting the possibility for 3D trapping. To confirm it three typical sizes ratios with negative axial and lateral radiation forces are selected for the three materials and the evolution of the force along $x$ and $z$ axes is studied. The axial radiation forces versus the position $z$ on the beam axis are given in the second row of Fig. \ref{Fig7: ASM less dense particles} for (d) PDMS, (e) olive oil, and (f) benzene, respectively, which show the ability of axial trapping. To further study the possibility of 3D trapping, the lateral force versus spatial position $x$ at the axial equilibrium position obtained in (d,e,f,) are plotted in the third row. The results confirm 3D trapping capabilities of (i) PDMS spheres with the three selected ratios $a/\lambda = $0.46, 0.83, and 1.35, (ii) olive oil sphere with $a/\lambda = $0.43, and (iii) benzene spheres with $a/\lambda = $0.39 and 1.06.  Note nevertheless that for PDMS the axial trap is asymmetric and much weaker than the lateral force and that this trend is further accentuated when inelasting scattering (absorption by the particle) is considered (see Appendix B). This is due to the fact that PDMS strongly absorbs the wave leading to scattering forces which push the particle in the wave propagation direction. So far, only levitation of these particle in the regime $a/\lambda \in [0.26 \, 0.52]$ has been reported \cite{cai2021self} and 3D trapping of PDMS particles requires further experimental confirmation.

\subsubsection{Conclusion}
The results of this subsection suggest the existence of 3D radiation trap for PDMS particles and Olive Oil and Benezene droplets at the center of one-sided focused beams both in the Rayleigh regime and beyond Rayleigh regime at specific frequencies.

%-----------------------Sec III
\subsection{\label{sec3C:3D cells trapping}  Biological particles}

The manipulation of biological microorganisms such as cells is of primary interest for applications. Here we review the abilities of focused beams to trap two types of biological particles, namely typical human cells and lipid cells.

\begin{figure*} [!htbp]
\includegraphics[width=17.6cm]{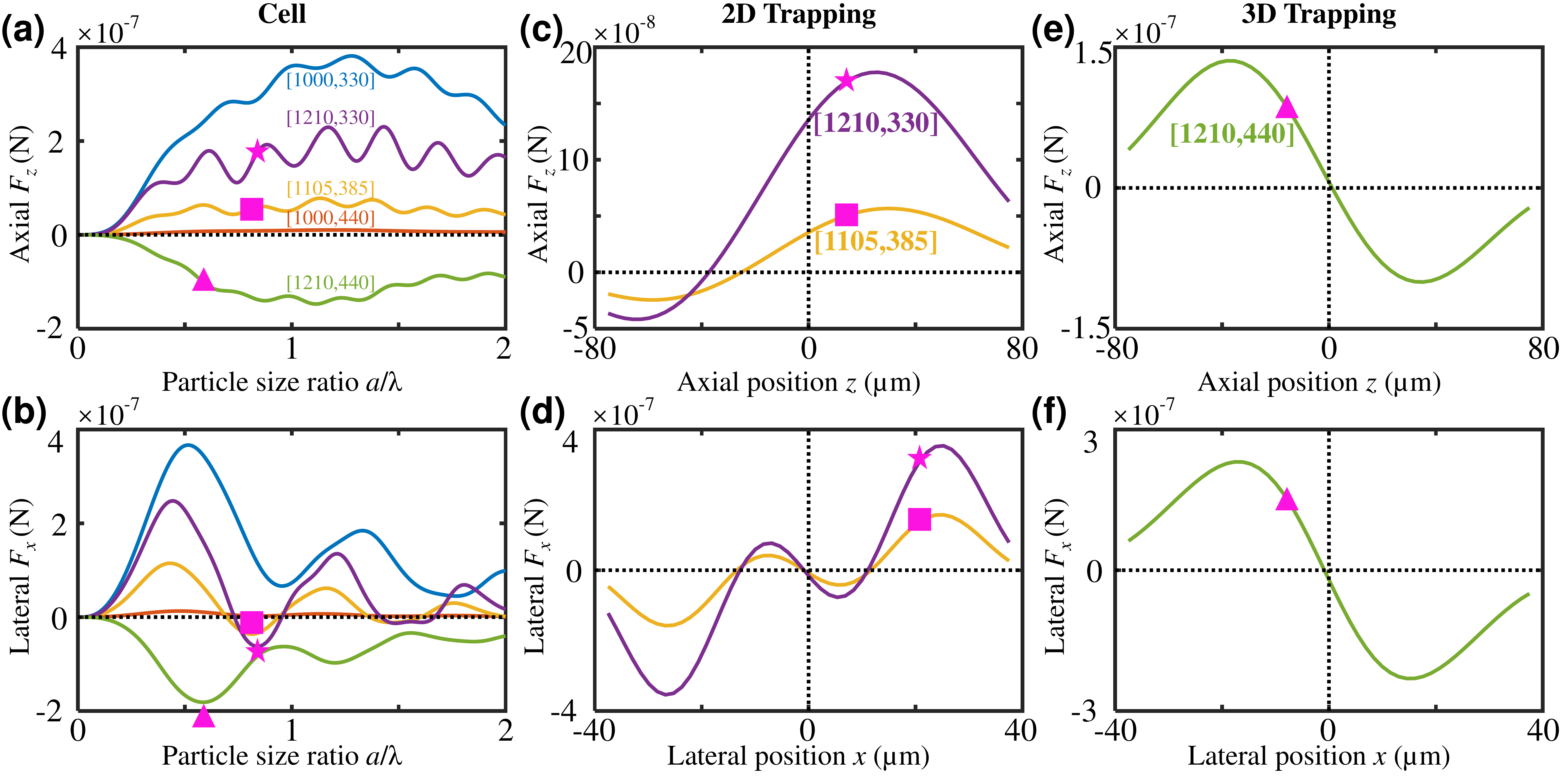}
\caption{Three-dimensional acoustical radiation forces based on the angular spectrum method at a fixed axial ($z_s = 30 \, \mu$m ) and lateral ($x_s = 8 \, \mu$m ) position for cells with different acoustic parameters of density (kg/m$^3$) and compressibility (1/TPa) [$\rho, \kappa$] = [1000, 330], [1000,440], [1105,385],[1210,330], and [1210,440]. (a) Axial force $F_z$ versus particle size ratio $a/ \lambda$, and (b) lateral force $F_x$ versus $a/ \lambda$. 
(c,d) give the axial and lateral radiation force versus positions for a fixed particle size as marked out in (b): $a/ \lambda = 0.81$ by \textcolor{magenta}{$\blacksquare$} for [$\rho, \kappa$] = [1105, 385] and $a/ \lambda = 0.84$ by \textcolor{magenta}{$\bigstar$} for [$\rho, \kappa$] = [1210, 330]. Only the lateral 2D trapping is possible. 
Similar to (c,d), (e,f) give the axial and lateral three-dimensional radiation force for a cell with $a/ \lambda = 0.58$ by \textcolor{magenta}{$\blacktriangle$} for [$\rho, \kappa$] = [1210, 440]. The lateral radiation force versus $x$ are plotted at the axial equilibrium positions in (f). At this case, a 3D trapping occurs.}
\label{Fig8: Cells for 2D and 3D trapping}
\end{figure*}

\subsubsection{\label{sec C1: cell}Typical human cells}
The density and and compressibility of typical human cells are taken from Ref. \cite{augustsson2016iso} and given in Table \ref{Table 1 Acoustic properties}. Typical cells are slightly less compressible and denser than water. Five sets of acoustic parameters corresponding to the extreme and average values of Ref. \cite{augustsson2016iso} are considered and the respective axial and lateral radiation forces versus size ratio $a/\lambda$ are given in Fig. \ref{Fig8: Cells for 2D and 3D trapping}(a) and (b). Note that these specific values do not correspond to specific types of cell but are used to give tendencies depending on the variation of the compressibility and density. As in the previous sections, the force is first calculated at a fixed axial position $z_s = 30 \, \mu$m and fixed lateral position $x_s = 8 \, \mu$m. As observed, the negative axial radiation force only occurs for $[\rho, \kappa]$ = [1210, 440], while the negative lateral force is possible for $[\rho, \kappa]$ = [1000, 440], [1105, 385], [1210, 330], and [1210, 440] at certain size over wavelength ratios. This first calculation suggest the possibility for 2D or 3D trapping for human cells with some acoustic parameters in a spherical focused beam. The examples of 2D trapping are given in Fig. \ref{Fig8: Cells for 2D and 3D trapping}(c) and (d), which show clearly the lateral trapping without axial trapping for the cases $[\rho, \kappa]$ = [1105, 385] with $a/\lambda=$ 0.81 and $[\rho, \kappa]$ = [1210, 330] with $a/\lambda=$ 0.84. 3D trapping only occurs for the the largest density and compressibility $[\rho, \kappa]$ = [1210, 440] compared with water as shown in Fig. \ref{Fig8: Cells for 2D and 3D trapping}(e) and (f). The axial force and lateral force at axial equilibrim position versus spatial positions are calculated for a size ratio $a/\lambda=$ 0.58. However these extreme values might not correspond to existing cells. Typical human cells could however be trapped by using single beam tweezers based on vortex beam \cite{gong2021Ztrap}.

\subsubsection{\label{sec C2: lipid}Lipid (fat) cells}
\begin{figure*} [!htbp]
\includegraphics[width=17.6cm]{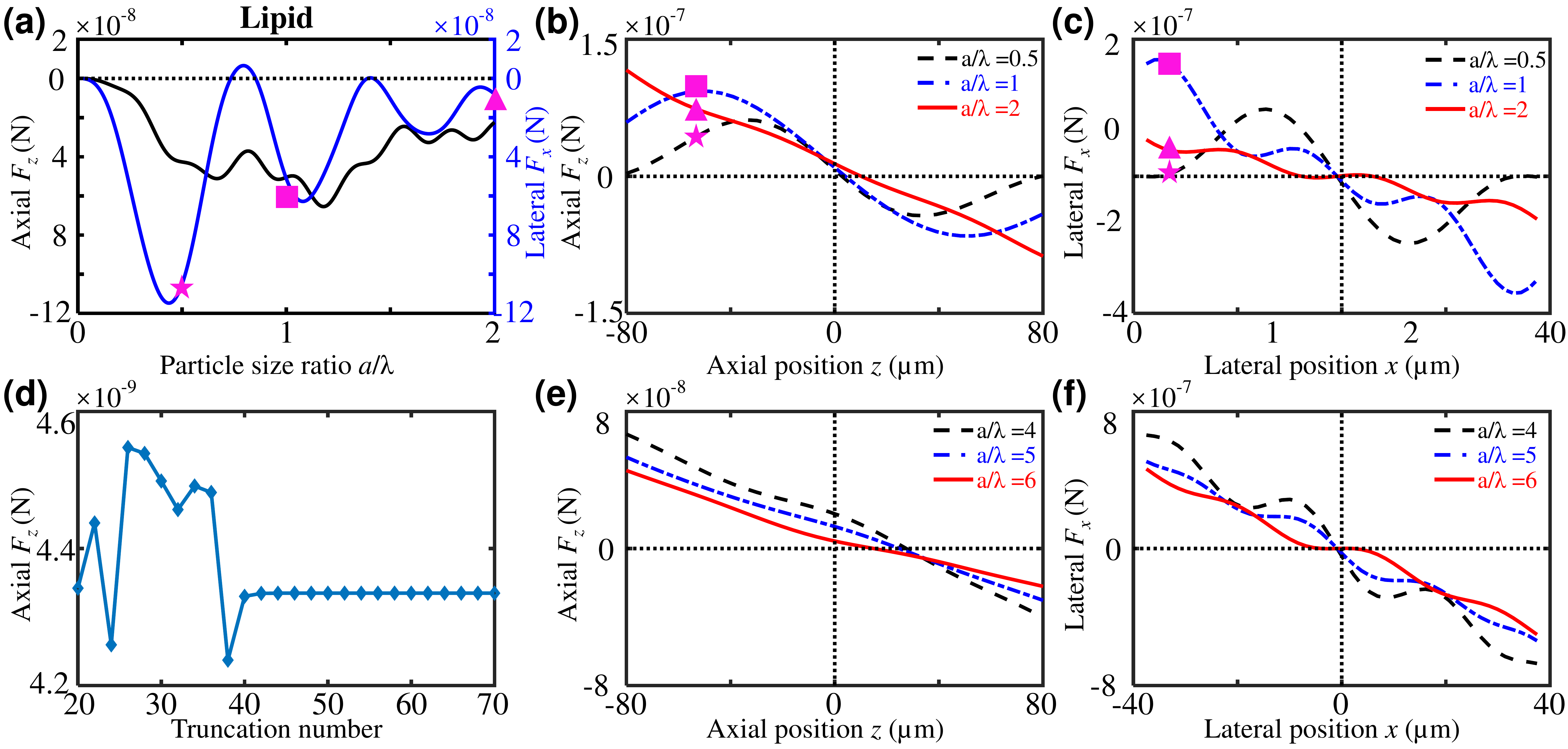}
\caption{(a) Three-dimensional acoustical radiation forces based on the angular spectrum method at a fixed axial ($z_s = 30 \mu$m ) and lateral ($x_s = 8 \mu$m ) position for lipid (fat cell) with density $\rho=$ 950 (kg/m$^3$) and sound speed $c=$ 1450 m/s. 
The left and right vertical axes are respective the axial ($F_z$, black solid line) and lateral ($F_x$, blue solid line) radiation force, while the horizontal axis is the ratio of particle radius over the wavelength $a/\lambda$ from 0 to 2.
(b,c) give the axial and lateral radiation force versus positions for a fixed particle size as marked out in (a) with $a/ \lambda = 0.5$ by \textcolor{magenta}{$\bigstar$}, $a/ \lambda = 1$ by \textcolor{magenta}{$\blacksquare$}, and $a/ \lambda = 2$ by \textcolor{magenta}{$\blacktriangle$}.
(d) shows a convergence test for high-frequency lipid cell with $a/ \lambda = 6$.
(e,f) give the axial and lateral radiation force versus positions for a fixed particle size with $a/ \lambda = 4$ , $a/ \lambda = 5$, and $a/ \lambda = 6$. The 3D trapping of lipid cells are possible in the designed focused bean at certain size ratios.}
\label{Fig9: lipid (fat) for 2D and 3D trapping}
\end{figure*}

The first experiments of single beam trapping with acoustical tweezers based on focused beams were conducted by Lee \textit{et al.} for the oleic acid lipid droplets and in this work only lateral 2D trapping was reported. The typical density of lipid cells are [910-1010] kg/m$^3$ \cite{neurohr2020relevance}. Here, we take the acoustic parameters (density and sound speed) of lipid droplet from Ref. \cite{lee2006radiation} as listed in Table \ref{Table 1 Acoustic properties}. 
At the fixed axial position $z_s = 30 \, \mu$m and the lateral position $x_s = 8 \, \mu$m, the 3D radiation forces versus the size ratios in the designed focused beam are first studied as given in Fig. \ref{Fig9: lipid (fat) for 2D and 3D trapping}(a), which suggest the possibility for 3D trapping in the range of  $a/\lambda=$ 0 to 2. This is further confirmed by Fig. \ref{Fig9: lipid (fat) for 2D and 3D trapping} (b) and (c) which show the axial and lateral radiation force versus the spatial position for three selected size ratios: $a/\lambda=$ 0.5, 1, and 2. The lateral forces in Fig. \ref{Fig9: lipid (fat) for 2D and 3D trapping}(c) are calculated at the axial equilibrium position obtained in (b). These figure show both an axial and lateral restoring force. We further investigate the trapping ability for the size rations used in Ref. \cite{lee2006radiation}, i.e. $a/\lambda=$ 4, 5, and 6. Since the truncation number in the angular spectrum method depends on the frequency, the computational cost is large when the particle size is in this regime, which make it hardly possible to calculate the 3D radiation forces versus size ratio like Fig. \ref{Fig9: lipid (fat) for 2D and 3D trapping}(a) with our simulation hardware platform with reasonable computation time. This is why we computed the radiation force only for these specific values and made some convergence tests as a function of the truncation number (Fig. \ref{Fig9: lipid (fat) for 2D and 3D trapping}(d)) for the worst case  $a/\lambda=$ 6. In this case, the truncation number \cite{gong2019reversals} $N_{max} = 2+ \text{Int}(8+ka+4.05 \sqrt[3]{k a})=62$ for Eq. (\ref{ASM force}) is larger than 42, which is the number starting to be convergent in Fig. \ref{Fig9: lipid (fat) for 2D and 3D trapping}(d), with $k$ the wavenumber and ``Int" denoting the integer part of the indicated argument. For the 3 calculated ratios $a/\lambda=$ 4, 5, and 6, the axial radiation force versus $z$ and the lateral forces versus $x$ at axial equilibrium position are given in (e) and (f), respectively. These figures suggest that 3D trapping of lipid cell at these particle size over wavelength ratio is possible.

\subsubsection{Conclusion}

Our results suggest (i) the possibility to trap typical human cells in 2D, while 3D trapping might be possible only for specific cells with large density and compressibility, and (ii) to trap lipid cell in 3D in Mie and geometric optics regimes at specific frequencies. 

%-----------------------Sec IV
\section{\label{sec4:conclusion} Conclusions}

Compared to their focused vortex counterpart, single beam acoustical tweezers based on focused beams have several advantages, such as easier synthesis, higher expected selectivity and forces due to stronger gradients, no repulsive ring surrounding the trapped particle, which complexifies particle assembly \cite{gong2019particle,gong2020three} and no rotation of the particle due to angular momentum transfer \cite{gong2020acousticart}. But so far, 3D trapping with focused beams has never been demonstrated. Our numerical analysis shows that single beam acoustical tweezers based on focused beams may have the potential (i) to trap elastic particles and droplets more compressible than the surrounding medium in 3D in and beyond Rayleigh regime; (ii) to trap less compressible particles in lateral 2D direction for some size ratios near resonances beyond Rayleigh regime; (iii) to trap lipid cells in 3D and typical human cells in 2D. This work provides a basis toward experimental investigation of 3D trapping abilities of droplets, particles and microorganisms with single beam tweezers based on focused beams. Next step would include experimental confirmation and also calculation and measurement of the streaming produced by the focused beam depending on the actuation frequency and comparison of the streaming induced drag force to the trapping force.

%-----------------------
\begin{acknowledgments}
We acknowledge the support of the programs ERC Generator, Prematuration, and Talent project funded by ISITE Universit\'{e} Lille Nord-Europe. We also thank Prof. Philip L. Marston at Washington State University in the U.S.A. for helpful discussion on the seeking for resonance frequencies of fluid sphere.
\end{acknowledgments}

%-----------------------
\appendix

\section{\label{Appendix A} Scattering coefficients of a sphere with different materials}
The scattering coefficients of sphere are well known in the literature and are reviewed in the Appendix of Gong's thesis \cite{gong2018thesis} which are recalled hereafter for convenience. Below, $k$ corresponds to the wavenumber in the fluid medium, $a$ to the radius of the sphere, $\rho$ to the density and  $c$ to the sound speed at rest, respectively.
\subsection{\label{Appendix A1} Rigid sphere}
As a background scattering to isolate the resonance contribution of an elastic sphere from the total scattering field, the scattering coefficients of a rigid sphere are 
\begin{equation}
s_{n}=-{h_{n}^{(2)}}'(k a) / {h_{n}^{(1)}} '(k a), 
\label{Rigid sphere}
\end{equation}
where the indexes (1) and (2) indicate the first and second kind of Hankel functions, and the prime $(')$ represents the derivative with respect to the indicated argument to the  argument $(ka)$.

\subsection{\label{Appendix A2} Soft sphere}
The scattering from a soft sphere can be considered as the background contribution of a bubble in liquid with the impedance smaller than that of the surrounding medium.
The scattering coefficients of a soft sphere are

\begin{equation}
s_{n}=-h_{n}^{(2)}(k a) / h_{n}^{(1)}(k a), 
\label{Soft sphere}
\end{equation}

\subsection{\label{Appendix A3} Fluid (liquid and air) sphere}
In an ideal fluid sphere, only the longitudinal wave exists (no transverse wave). A parameter $D_n$ is introduced for convenience with the relation to the scattering coefficients given by $s_n = -{D_n^*}/{D_n}$ \cite{marston2006axial}, where

\begin{equation}
D_{n}=\rho_{f} k a j_{n}\left(k a / \gamma_{c}\right) {h_{n}^{(1)}}' (k a)-\rho\left(k a / \gamma_{c}\right) {j_{n}}^{\prime}\left(k a / \gamma_{c}\right) h_{n}^{(1)}(k a), 
\label{fluid sphere}
\end{equation}
and the asterisk $^*$ indicates the complex conjugate,  $\rho_{f}$ the density of the fluid sphere, $\gamma_{c} = c_f / c$ the ratio of sound speed in the spherical fluid droplet ($c_f$) over that in the surrounding fluid $c$, and $j_n$ the Bessel function of the first kind.

\subsection{\label{Appendix A4} Elastic sphere}
For an elastic sphere, there are both longitudinal and transverse components of elastic waves with their sound speed $c_l$ and $c_t$, and wavenumber $k_l = (c/c_l)k$ and $k_t = (c/c_t)k$, respectively. The density of the elastic sphere is $\rho_e$. It is convenient to define the dimensionless frequency in the fluid medium of longitudinal wave in the elastic sphere  $x_l = k_l a$, and of transverse wave $x_t = k_t a$. A coefficient is introduced as $N = n(n+1)$ for convenience. The scattering coefficients can be obtained by $s_n = -{ \left| D_n^*  \right|}/{\left| D_n \right|}$ with $D_n$ consisting of $3 \times 3$ elements \cite{gaunaurd1983rst}

\begin{align}
\begin{split}
% first vortex
&d_{11}=\left(\rho / \rho_{e}\right) x_{s}^{2} h_{n}^{(1)}(x) \\
&d_{12}=\left(2 N-x_{s}^{2}\right) j_{n}\left(x_{p}\right)-4 x_{p} {j_{n}}'\left(x_{p}\right) \\
&d_{13}=2 N\left[x_{s} {j_{n}}'\left(x_{s}\right)-j_{n}\left(x_{s}\right)\right] \\
&d_{21}=-{x h_{n}^{(1)}}'(x) \\
&d_{22}=x_{p} {j_{n}}'\left(x_{p}\right) \\
&d_{23}=N j_{n}\left(x_{s}\right) \\
&d_{31}=0 \\
&d_{32}=2\left[j_{n}\left(x_{p}\right)-x_{p} j_{n}^{\prime}\left(x_{p}\right)\right] \\
&d_{33}=2 x_{s} {j_{n}}'\left(x_{s}\right)+\left(x_{s}^{2}-2 N+2\right) j_{n}\left(x_{s}\right) .
\label{elastic sphere}
\end{split}
\end{align}
where the symbol `$\left|   \right|$' is the determinant of the matrices $D_n^*$ and $D_n$. Note that for an elastic shell, the explicit elements of $D_n$ are given in Ref. \cite{marston1989observations}.

\begin{figure} [!htbp]
\includegraphics[width=1\linewidth]{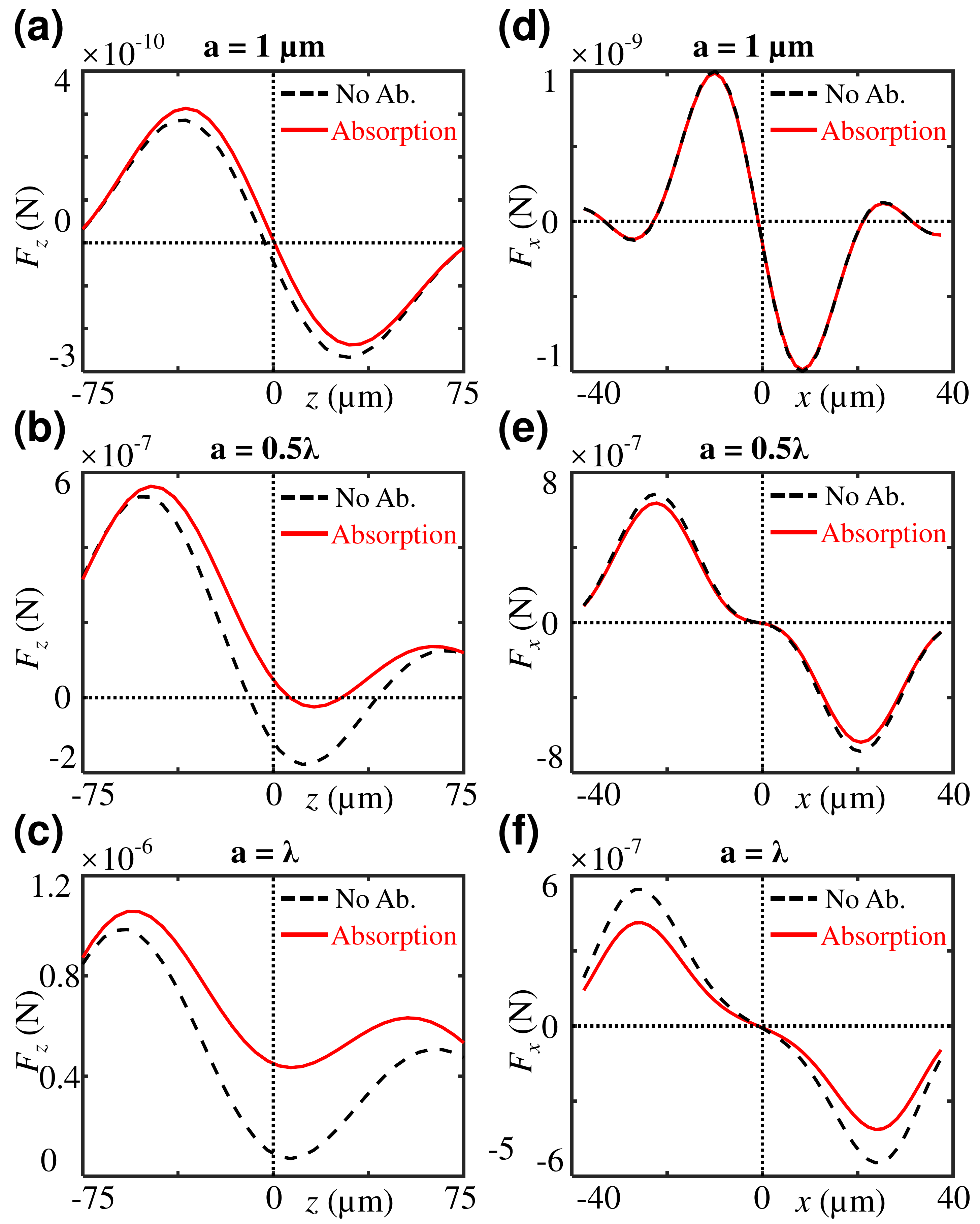}
\caption{The three-dimensional radiation forces versus the spatial positions for the PDMS particles with and without absorption in the axial (a,\, b,\, c) and lateral (d,\, e,\, f at the focal plane $z_s = 0$) directions. Three particle radii are taken into consideration as $a=1 \, \mu$m, $0.5\lambda$, and $\lambda$. The wavelength is $\lambda= 37.5 \, \mu$m.}
\label{Fig11: PDMS Absorption}
\end{figure}

\subsection{\label{Appendix A5} Viscoelastic spherical shell filled with fluid}
The scattering coefficients $s_n$ of a viscoleastic shell filled with fluid can be obtained based on the Kelvin-Voigt linear viscoelastic model \cite{gaunaurd1978theory,gong2020acousticart}, and calculated from the partial wave coefficients $A_n$ with the relation $A_n = (s_n -1)/2 $, where
\begin{equation}
A_{n}=-\frac{F_{n} j_{n}(x)-x j_{n}{ }^{\prime}(x)}{F_{n} h_{n}^{(1)}(x)-x h_{n}^{(1)}{} ^{\prime} (x)}, 
\label{VE sphere}
\end{equation}
Here, the time harmonics $e^{-i \omega t}$ is applied so that the Hankel function of the first kind should be used.
The explicit elements of $F_n$ are given in detail in Appendix A of Ref. \cite{gong2019reversals}. The complex wavenumber is used when the absorption is considered for the viscoelastic material (i.e., the contribution from the imaginary part). When the imaginary part vanishes, the model turns to an elastic material. In addition, the shell model can degenerate into a solid elastic sphere when the inner fluid is missing.

%\subsection{\label{Appendix A1}  xxx}

%-----------------------
\section{\label{Appendix B PDMS with absorption} PDMS in viscoelastic model}

The Kelvin-Voigt linear viscoelastic model \cite{gaunaurd1978theory,gong2020acousticart} is applied to calculate the scattering coefficients as given in Appendix \ref{Appendix A5} when the absorption effects are considered inside the PDMS particles.
The normalized absorption coefficients of the PDMS are calculated from Ref. \cite{guo2020acoustic}. Here, the normalized absorption coefficients of the longitudinal and transverse waves are $\gamma_l=0.0075$ and $\gamma_t=0.2673$, respectively.
The 3D radiation forces for PDMS particles with and without absorption are studied based on the angular spectrum method with radii $a= 1 \, \mu$m, $0.5\lambda$, and $\lambda$, as shown in Fig. \ref{Fig11: PDMS Absorption}.  
As shown, the 3D trapping is still possible for small particles when the absorption effect is under consideration, although the axial pushing force increases due to the fact that the absorption of the linear momentum produce the positive axial radiation force \cite{zhang2011geometrical}. When the particle size reaches $a= \lambda$, for instance, there is no axial trapping anymore while the lateral trapping is still possible, see Fig. \ref{Fig11: PDMS Absorption}(c) and (f). This gives us some guidance to use a focused beam to trap compressible elastic particles in three dimensions for experimental demonstration, which has not yet been done before to the authors' knowledge. It is noteworthy that the levitation of PDMS particles in Mie regime ($a/\lambda \approx 1$) by a focused beam using transducer array has recently been implemented \cite{cai2021self}. However, the 3D trapping was not demonstrated.

\renewcommand\refname{Reference}
\bibliography{main}        %Produce the bibliography

\end{document}